\documentclass[prb,twocolumn,amsmath,amssymb,floatfix,footinbib,bibnotes,longbibliography]{revtex4-1}
\pdfoutput=1

\usepackage[colorlinks=true,citecolor=blue,linkcolor=magenta]{hyperref}
\usepackage{amsmath}
\usepackage{graphicx}
\usepackage{changepage}
\usepackage{fancyhdr}
\usepackage{amsthm, amssymb}
\usepackage{floatflt}
\usepackage{float}
\usepackage{array}
\usepackage[usenames,dvipsnames]{color}
\usepackage[section]{placeins}
\def \mr{\mathrm}

\def \mc{\mathcal}

\newcommand{\ci}{\mathfrak{i}}
\newcommand {\apgt} {\ {\raise-.5ex\hbox{$\buildrel>\over\sim$}}\ }
\newcommand {\aplt} {\ {\raise-.5ex\hbox{$\buildrel<\over\sim$}}\ }

\def \titlename {Entanglement in non-equilibrium steady states and many-body localization breakdown in a current driven system}
\def \authornames{Animesh Panda, and Sumilan Banerjee}
\def \affiliations{Centre for Condensed Matter Theory, Department of Physics, Indian Institute 
of Science, Bangalore 560012, India}
\begin{document}

\title{\titlename}
\author{\authornames}
\affiliation{\affiliations}

\date\today

\begin{abstract}
 We model a one-dimensional (1D)  current-driven interacting disordered system through a non-Hermitian Hamiltonian with asymmetric hopping and study the entanglement properties of its eigenstates. In particular, we investigate whether a many-body localizable system undergoes a transition to a current-carrying non-equilibrium steady state under the drive and how the entanglement properties of the quantum states change across the transition.  We also discuss the dynamics, entanglement growth, and long-time fate of a generic initial state under an appropriate time-evolution of the system governed by the non-Hermitian Hamiltonian. Our study reveals rich entanglement structures of the eigenstates of the non-Hermitian Hamiltonian. We find transition between current-carrying states with volume-law to area-law entanglement entropy, as a function of disorder and the strength of the non-Hermitian term. 
\end{abstract}

\maketitle
\section{Introduction}

 Classification of many-body quantum states in terms of their entanglement properties has become a major theme in condensed matter physics. These activities have revealed intriguing connections of entanglement with dynamics and thermalization of quantum systems \cite{Altman2015,Nandkishore2015,Abanin2018}. 
 It has been understood that typical high-energy eigenstates of a generic \emph{isolated} many-body system can be classified as either \emph{ergodic} or \emph{non-ergodic} depending on the subsytem entanglement entropy, e.g. $S_{EE}=-\mr{Tr}_A\rho_A\ln\rho_A$ of a subsytem $A$, where $\rho_A$ is the reduced density matrix of $A$ for the eigenstate. In particular, when $d$-dimensional system is divided into two subsytems with the smaller subsystem, say $A$, having length $L$, the entanglement entropy $S_{EE}\sim L^d$, i.e. scales with the volume of the subsystem. On the contrary, for \emph{many-body localized} (MBL) states \cite{Gornyi2005,Basko2006,Oganesyan2007,Pal2010,Altman2015,Nandkishore2015,Abanin2018}, prominent examples of non-ergodic eigenstate, subsytem entanglement scales as $L^{d-1}$, i.e. with the \emph{area} of the subsystem, the so-called \emph{area law} scaling. Starting with a generic initial state, the MBL systems do not thermalize under the unitary dynamics, unlike the ergodic ones. Nevertheless, the MBL systems still give rise to a slow growth of entanglement entropy approaching a state with sub-thermal volume law scaling \cite{Znidaric2008,Bardarson2012,Serbyn2013,Vosk2013}. One of the most natural realizations of MBL phases are found in strongly disordered systems where single-particle Anderson localization is stable to interaction \cite{Basko2006,Gornyi2005,Oganesyan2007,Pal2010}, even at finite energy densities above the ground state.  A many-body localizable system can undergo a volume-law (ergodic) to area-law (non-ergodic) transition \cite{Gornyi2005,Basko2006,Oganesyan2007,Pal2010,Altman2015,Nandkishore2015,Abanin2018} as a function of the strength of the disorder or the interaction .
 
\begin{figure}[htb!]
	\centering
	{
		\includegraphics[width=0.4\textwidth]{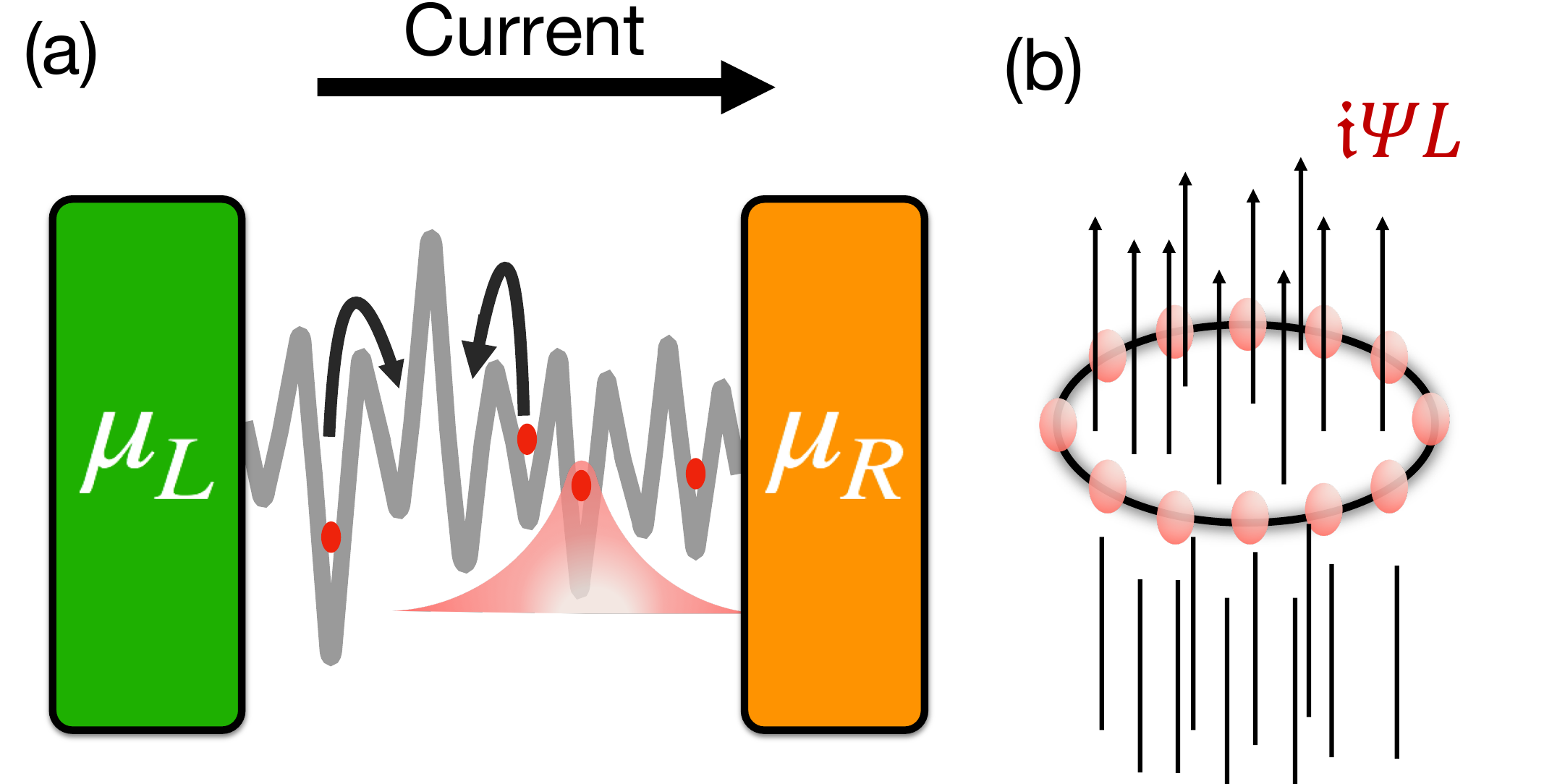}
	}
	
	\caption{{\bf The model:} (a) Schematic of an interacting disordered system driven by current through a voltage bias, $\mu_\mr{R}-\mu_\mr{L}$, applied between left and right leads with chemical potentials $\mu_\mr{L}$ and $\mu_\mr{R}$, respectively. (b) The non-Hermitian model [Eq.\eqref{eq:FermionModel}] for interacting fermions hopping on a ring with an imaginary flux $i\Psi L$.}
	\label{fig:Model} 
\end{figure}

 However, condensed matter systems are seldom isolated. In particular, some very important experimental setups require the system to be connected with external environment. A notable example is a system connected to leads and driven by current or a voltage bias or an electric field [see Fig.\ref{fig:Model}(a)]. Generically, such systems are expected to attain an unique current-carrying non-equilibrium steady state (NESS). It is interesting to explore whether such NESSs could also be classified according to their entanglement content. For example, we could ask whether  a many-body localizable system can undergo a dynamical transition when driven by a strong current or electric field, e.g. a transition between NESSs with distinct system-size scaling of current. Moreover, can such NESS transition be coincident with an entanglement transition?
 
  Recently there have been some studies of entanglement properties \cite{Sharma2015,Dey2019} as well as entanglement transitions \cite{Gullans2018,Gullans2019} for the current carrying NESSs in non-interacting models using non-equilibrium Green's functions or scattering states. However, computing entanglement properties of driven states of a disordered interacting system connected to infinite leads at the boundaries is an extremely challenging task. One possible way to access such boundary driven system is via Markovian Lindblad quantum master equation approximation \cite{Prosen2009,Znidaric2014,Mahajan2016,Zanoci2016}. Nevertheless, generically, such Markovian evolution is destined to lead to a description of NESS in terms of mixed state having area law entanglement, e.g. quantified in terms of mutual information \cite{Wolf2008,Mahajan2016,Zanoci2016}, and hence make the notion of entanglement transition obscure.  Also, in non-interacting systems, there are examples of NESSs with higher than area-law mutual information \cite{Eisler2014}, or even volume-law entanglement \cite{Gullans2018,Gullans2019} . Such highly entangled NESSs can persist even in interacting systems with size smaller than electronic phase coherence length \cite{Gullans2018}. An important avenue, at present, is to try to mimic such NESS in less microscopic \emph{toy} models, e.g. boundary driven random unitary circuit model \cite{Gullans2018}. Here we take a different route, and try to address driven states of interacting disordered systems through an effective model which incorporates the dissipation and the current drive via a non-Hermitian term.
 
 The model studied here is an interacting version of the one dimensional Hatano-Nelson model \cite{Hatano1996,Hatano1997,Hatano1998} for non-Hermitian single-particle localization-delocalization transition. The model has a time ($\mc{T}$)-reversal symmetry, that could be spontaneously broken by the eigenstates for a large enough strength of the non-Hermitian term. In the non-interacting Hatano-Nelson model, $\mc{T}$-reversal symmetry breaking or real-to-complex eigenvalue transition \cite{Hatano1996,Hatano1997,Hatano1998} coincides with localization-delocalization transition. We study the interacting non-Hermitian model via exact diagonalization (ED). Our main results are the following.\\
 (1) We show that a current-driven many-body localizable system can undergo a volume- to area-law entanglement transition in the eigenstates as a function of disorder and/or the strength of the non-Hermitian term for a fixed interaction strength.\\
 (2) We find that entanglement transition has a direct correspondence with a transition, similar to MBL transition in Hermitian system \cite{Luitz2015,Mace2019}, in terms of participation of the eigenstates in the many-body Hilbert space.\\ (3) We demonstrate that the entanglement and $\mc{T}$-reversal breaking transitions are distinct in the interacting case. We also find another distinct transition within the $\mc{T}$-reversal broken region in terms of the scaling of the current with system length.\\
(4) We show how the system approaches the NESS at long times, under an appropriately defined dynamics, in terms of time-evolution of the entanglement entropy and the memory of the initial state. In particular, as in the Hermitian MBL systems, starting with an initial unentangled (product) state the entanglement entropy grows linearly with time $t$ in the volume-law phase. In contrast, both in the $\mc{T}$-reversal broken and unbroken area-law phase, entanglement entropy grows as $\ln(t)$. However, the entanglement growth is followed by a decay at late times towards an unique NESS. Moreover, we find that the memory of the initial state always eventually gets lost as the NESS is attained at long times.
\\
Overall, our study reveals a much richer eigenstate and dynamical phase diagram of the non-Hermitian model compared to the Hermitian systems.

 The remainder of this paper is organized as follows. In Section~\ref{sec:Model} we
describe the model and the dynamics governed by the non-Hermitian Hamiltonian. We give some physical motivations behind the model and its dynamics in Section \ref{sec:motivation}.
Section~\ref{sec:Results} discusses the results for
the Hilbert-space localization-delocalization, entanglement and time-reversal symmetry breaking transitions in terms of finite-size scaling analysis. We also describe here the transient dynamics of the system during the approach to NESS and the properties of the NESS. In
Sec.~\ref{sec:Conclusions} we conclude with the implications and significance of our results. The appendix include some additional results.
\section{Model} \label{sec:Model}
In this section we introduce the non-Hermitian Hamiltonian and the dynamics. In the next section, we briefly review general motivations for the model, based on non-Hermitian approach to open systems, and discuss some specific physical realizations.

\subsection{Non-Hermitian Hamiltonian}
 We study the following non-Hermitian one-dimensional ($1D$) $XXZ$ spin ($S=1/2$) model with an uniform random field $h_i\in [-W,W]$, $W$ being the disorder strength,
 \begin{align}
 \mc{H}&=-\frac{J}{2}\sum_i^L(e^\Psi S_i^+S_{i+1}^-+e^{-\Psi}S_i^-S_{i+1}^+)-\sum_i^L h_iS_i^z\nonumber \\
 &-J_z\sum_i^L S_i^zS_{i+1}^z. \label{eq.SpinModel}
 \end{align}  
 Here $S_i^\pm=(S_i^x\pm iS_i^y)$, $L$ is the number of sites, $\Psi$ is a real number (see below) and  $J,~J_z$ are the spin exchanges; the latter controls the interaction strength. We apply periodic boundary condition. The above model can be rewritten as $\mc{H}=\mc{H}_h+\ci\lambda \mc{J}$, where $\mc{H}_h$ is the usual Hermitian random field $XXZ$ model with $J\to\tilde{J}=J\cosh{\Psi}$, $\lambda=J\sinh{\Psi}$ and $\mc{J}=-(\ci J/2)\sum_i (S_i^+S_{i+1}^--S_i^-S_{i+1}^+)$ is the sum of spin current across the system. 
 
 The model can be mapped, via Jordan-Wigner transformation, to a model of spin-less fermions hopping (with amplitude $t=J/2$) on the $1D$ lattice with random disorder potential ($\epsilon_i=h_i$) and nearest neighbor interaction ($V=J_z$), i.e.
\begin{align}
\mc{H}&=-t\sum_i^L(e^\Psi c_i^\dagger c_{i+1}+e^{-\Psi}c_{i+1}^\dagger c_i)-\sum_i^L\epsilon_i n_i \nonumber \\
&-V\sum_i^Ln_i n_{i+1}. \label{eq:FermionModel}
\end{align}
Here $c_i,c_i^\dagger$ are fermion operators. In this case, $\ci\Psi$ is an imaginary vector potential corresponding to an imaginary flux $\ci\Psi L$ through a ring of circumference $L$ [Fig.\ref{fig:Model}(b)] and the model is an interacting version of Hatano-Nelson model \cite{Hatano1996,Hatano1997,Hatano1998}. The latter describes non-Hermitian single-particle localization-delocalization transition. The spinless fermion model is invariant under $\ci=\sqrt{-1}\to-\ci$, i.e. time ($\mc{T}$) reversal. However, as in the usual $\mathcal{P T}$-symmetric non-Hermitian models \cite{Bender1998,Bender2007}, this $\mc{T}$- or pseudo $\mathcal{P T}$-symmetry can be broken by the eigenstates leading to complex eigenvalues. We refer to this real to complex transition as $\mc{T}$-reversal breaking for brevity even in the spin model [Eq.\eqref{eq.SpinModel}]. Intriguingly, in the original Hatano-Nelson model the $\mc{T}$-symmetry breaking transition of the single-particle eigenstates coincides with the localization-delocalization transition and the delocalized states carry finite current. This leads to the question whether there is any \emph{localization-delocalization} transition in the interacting model and whether the transition coincides with the $\mc{T}$-reversal breaking. Such congruence of the symmetry breaking and localization transition might lead to an exciting possibility of describing a ``MBL transition" in terms of symmetry breaking.

\subsection{Non-Hermitian dynamics}
We further extend the model of Eq.\eqref{eq.SpinModel} to describe the approach to the long-time NESS through the following dynamical equation for density matrix $\rho$ ($\hbar=1$),
\begin{align}
\frac{d\rho}{dt}=-\ci[\mc{H}_h,\rho]+\lambda\left(\{\mc{J},\rho\}-2\mr{Tr}(\rho \mc{J})\rho\right)\equiv \mc{L}\rho. \label{eq.DynEq}
\end{align}
Here $\mc{L}$ defines a Liouvillian operator. As we discuss in the next section, the above dynamics for the density matrix can be obtained starting with the time-dependent non-Hermitian Schr$\ddot{\mr{o}}$dinger equation \cite{Sergi2013,Zloshchastiev2014}.
The dynamical evolution of Eq.\eqref{eq.DynEq} has been used previously in the context of $\mc{PT}$-symmetric quantum mechanics to describe system with gain and loss \cite{Brody2012}. A similar model has been also used in the context of field-driven Mott transition \cite{Tripathi2016}. Eq.\eqref{eq.DynEq} can describe the evolution of both pure and mixed states as discussed in Ref.\onlinecite{Brody2012} and keep $\mr{Tr}\rho(t)=1$. Unlike in a Lindblad master equation, an initial pure state remains pure during the time evolution, and, in this case, Eq.\eqref{eq.DynEq} reduces to a complex Schr$\ddot{\mr{o}}$dinger equation \cite{Brody2012}. From Eq.\eqref{eq.DynEq}, $\rho(t)$ can be formally written using the non-Hermitian Hamiltonian [Eq.\eqref{eq.SpinModel}] as $\rho(t)=e^{-\ci\mc{H}t}\rho_0e^{\ci\mc{H}^\dagger t}/\mr{Tr}(e^{-\ci\mc{H}t}\rho_0e^{\ci\mc{H}^\dagger t})$, where $\rho_0$ is the initial density matrix. In particular, for an initial pure state $\rho_0=|\psi\rangle \langle \psi |$, the state at time $t$ is given by 
\begin{align}
|\psi(t)\rangle&=\frac{\sum_n e^{-\ci \mc{E}_n t+ \Lambda_n t}\langle n_L|\psi\rangle |n_R\rangle}{\|\sum_n e^{-\ci\mc{E}_n t+\Lambda_n t}\langle n_L|\psi\rangle |n_R\rangle\|} \label{eq:purestate}
\end{align}
Here $\langle n_L|$ and $|n_R\rangle$ are the left and right eigenvectors of $\mc{H}$ with eigenvalue $E_n=\mc{E}_n+\ci \Lambda_n$; $\|.\|$ denotes the norm of a vector. As well known \cite{Faisal1981}, the left and right eigenvectors form a bi-orthonormal basis, $\langle n_L|m_R\rangle=\delta_{nm}$ with the resolution of the identity $\sum_n|n_R\rangle\langle n_L|=\mathbb{I}$. The real part of the eigenvalue $\mc{E}_n=\langle n_R|\mc{H}_h|n_R\rangle/\langle n_R|n_R\rangle$, i.e. the expectation of the parent Hermitian Hamiltonian, and the imaginary part $\Lambda_n\propto\mc{J}_n=\langle n_R|\mc{J}|n_R\rangle/\langle n_R|n_R\rangle$. Here the bra vector $\langle n_R|=|n_R\rangle^\dagger$ is the Hermitian conjugate of the right eigenket, and similarly $|n_L\rangle=(\langle n_L|)^\dagger$. The identifications of real part and imaginary parts of the eigenvalue as the eigenstate expectation value of $\mathcal{H}_h$ and $\mathcal{J}$ are due to the fact that  both the operators are Hermitian and hence their expectation values are real. It can be easily shown from Eq.\eqref{eq:purestate} that, for eigen spectrum with non-zero imaginary part of the eigenvalues,  the NESS $|\psi_\infty\rangle\equiv |\psi(t\to\infty)\rangle$ is given by 
\begin{align}
|\psi_\infty\rangle=\frac{\langle s_L|\psi\rangle}{|\langle s_L|\psi\rangle|}\frac{|s_R\rangle}{\| |s_R\rangle\|} \label{eq:NESS}
\end{align}
 where $\langle s_{L}| (|s_{R}\rangle )$ is the left (right) eigenvector with maximum imaginary part of the eigenvalue $\Lambda_s$. 

\section{Motivations and physical realizations}\label{sec:motivation}
\subsection{Non-Hermitian approach to open quantum systems}
Non-Hermitian Hamiltonians have been used in many past studies to describe decaying states (e.g. see Refs.\onlinecite{Feshbach1958,Feshbach1962,CohenTannoudji1968,Faisal1981}) and effects of dissipation and drive in open quantum systems (e.g. see Refs.\onlinecite{Sergi2013,Zloshchastiev2014,Rotter2009,Brody2012,Tripathi2016,Fukui1998,Oka2010}).
For the latter systems, the non-Hermitian approach has range of applicability different and complementary to more widely used Lindblad master equation approximation \cite{WeissBook}. For instance, as already discussed, the non-Hermitian approach could describe both pure and mixed states and their dynamical evolutions \cite{Brody2012,Sergi2013,Zloshchastiev2014}. The general theoretical framework, in principle, to rigorously derive such non-Hermitian Hamiltonian is rooted in the Feshbach projection operator technique \cite{Feshbach1958,Feshbach1962,CohenTannoudji1968,Faisal1981,Rotter2009}. The latter can be used to construct non-Hermitian Hamiltonian to describe the effects of continuum of scattering states in the presence of an environment on a finite subsystem which has discrete states in isolation. However, here, as in earlier studies \cite{Tripathi2016,Fukui1998,Oka2010}, we use the model of Eq.\eqref{eq.SpinModel} from a more phenomenological ground to describe NESS in a current driven interacting disordered open system. Apart from the connections to some of the possible physical realizations that we discuss below, the model of Eq.\eqref{eq.SpinModel} could be thought of as an effective model to generate and study interesting current carrying pure states which are relatively simpler to access within exact numerical calculations, albeit for finite systems of modest sizes. Also, in the spirit of study of current carrying states in quantum field theories (see e.g. Ref.\onlinecite{Cardy2000}), the non-Hermitian term $i\lambda \mathcal{J}$ can also be interpreted as a Lagrange multiplier term that acts as a source field for inducing finite current in the eigenstates.

Once the non-Hermitian Hamiltonian of Eq.\eqref{eq.SpinModel} is posited as an effective phenomenological model to describe current driven open system, the dynamical evolution [Eq.\eqref{eq.DynEq}] results naturally from the corresponding non-Hermitian time-dependent Schr$\ddot{\mr{o}}$dinger equation \cite{Faisal1981} as discussed in Ref.\onlinecite{Sergi2013,Zloshchastiev2014}. We briefly review here the derivation of Eq.\eqref{eq.DynEq} for the sake of completeness. 

For the non-Hermitian Hamiltonian the time-dependent Schr$\ddot{\mr{o}}$dinger equation, $\ci\partial|\tilde{\psi}(t)\rangle/\partial t=(\mc{H}_h+i\lambda \mc{J})|\tilde{\psi}(t)\rangle$ for a state $|\tilde{\psi}(t)\rangle$, leads to
\begin{align}
\frac{\partial \tilde{\rho}}{\partial t}&=-\ci[\mc{H}_h,\tilde{\rho}]+\lambda\{\mc{J},\tilde{\rho}\} \label{eq:DynUnnormRho}
\end{align}
for a density matrix $\tilde{\rho}(t)=|\tilde{\psi}(t)\rangle \langle \tilde{\psi}(t)|$. As discussed in Refs.\onlinecite{Sergi2013,Zloshchastiev2014}, the above evolution does not preserve $\mr{Tr}\tilde{\rho}(t)$. However, the expectation values of an operator can be obtained using the normalized density matrix $\rho(t)=\tilde{\rho}(t)/\mr{Tr}\tilde{\rho}(t)$. It is straightforward to obtain the time-evolution of Eq.\eqref{eq.DynEq} for the normalized density matrix $\rho$, and hence for the normalized pure state $|\psi(t)\rangle$ [Eq.\eqref{eq:purestate}] with $\rho(t)=|\psi(t)\rangle \langle \psi(t)|$, from Eq.\eqref{eq:DynUnnormRho}. Therefore, we use Eq.\eqref{eq.DynEq} to describe the dynamics in our model.

\subsection{Physical realizations}

{\bf 1. Electric field-driven breakdown in interacting disordered systems:}
One of the motivations for the models of Eq.\eqref{eq.SpinModel} and Eq.\eqref{eq:FermionModel} comes from earlier works in Refs.\onlinecite{Fukui1998,Oka2010,Tripathi2016} on the electric field driven Mott transition. In this case, the imaginary gauge potential in Eq.\eqref{eq:FermionModel} arises from 
a asymmetric hopping amplitudes, $t_{1L}=-te^{L\Psi}\neq t_{L1}=-te^{-L\Psi}$, between site 1 and $L$ on the ring [Fig.\ref{fig:Model}(b)], while all other hoppings between nearest-neighbor sites are $-t$. It was argued in Ref.\onlinecite{Fukui1998} that these asymmetric hoppings mimic the situation of an open chain, connected to leads at the two ends, where electron is supplied at site 1 and dissipated at site $L$ at the other boundary. Thus the asymmetric hoppings give rise to both dissipation and drive. The model of Eq.\eqref{eq:FermionModel} can then be easily obtained via non-unitary gauge transformations, $c_i^\dagger \to e^{-i\Psi}c_i^\dagger$ and $c_i\to e^{i\Psi} c_i$ at site $i$, and $\ci\Psi L$ acts as an imaginary flux through the ring [Fig.\ref{fig:Model}(b)].

 In the case of electric field-driven Mott localization-delocalization (insulator-to-metal transition), the model with imaginary gauge potential has been shown \cite{Oka2010} to describe many-body Landau-Zener (LZ) \cite{Landau1958} quantum tunneling processes near field driven Mott transition within Dykhne \cite{Dykhne1962} formalism. In the latter approach, a model with real gauge potential, such as for a constant electric field, is analytically continued to imaginary gauge potential to describe  LZ transition. 

Similar to the electric-field driven Mott transition in the clean fermionic Hubbard model \cite{Fukui1998,Oka2010,Tripathi2016}, the non-Hermitian model can be thought of as an effective model to describe possible electric-field driven transition in a MBL system through many-body Landau-Zener breakdown. As in the case of field-driven Mott transition \cite{Oka2003}, the connection of the non-Hermitian model with field-driven MBL breakdown can be explored by considering interacting fermions on a ring [Fig.\ref{fig:Model}(b)] under a constant electromotive force applied by a time-dependent (real) flux through the ring. It is indeed an interesting question whether such many-body LZ processes could provide an effective description for field or current driven transition in a MBL system. Here we do not attempt to address this issue, and, instead, use the non-Hermitian model [Eq.\eqref{eq.SpinModel}] as an effective model to generate current driven pure states and describe possible entanglement transitions among them.

{\bf 2. Anomalous boundary states of symmetry protected gapped topological (Hermitian) systems:} Recently, new correspondence has been established between topological classification of $(d+1)$ dimensional gapped Hermitian systems and $d$ dimensional non-Hermitian systems \cite{Lee2019}. For example, as discussed in Ref.\onlinecite{Lee2019}, the long-time steady state of non-interacting clean non-Hermitian model in Eq.\eqref{eq:FermionModel} with $W=0$ and $V=0$ describes 1D chiral fermions, that are naturally realized at the edge of quantum Hall systems. Hence, the model of Eq.\eqref{eq:FermionModel} could describe the effects of disorder and interaction on such chiral edge states. 

{\bf 3. Depinning transition in classical systems of interacting vortices:} The original motivation behind the Hatano-Nelson model \cite{Hatano1996,Hatano1997,Hatano1998} was to study a classical systems of superconducting vortices  with columnar defects and tilted magnetic field applied at an angle with columnar defects in one higher dimension than the quantum model. The latter can be mapped to the classical model via standard path integral approach, and the depinning transition is understood as the localization-delocalization transition in the quantum model. In the same spirit, the interacting model of Eq.\eqref{eq:FermionModel} may describe the effects of inter-vortex interaction on the depinning transition \cite{Hamazaki2019}.

\section{Results} \label{sec:Results}
\begin{figure*}[!htb]
	\centering{
		\includegraphics[width=\linewidth,height=45mm]{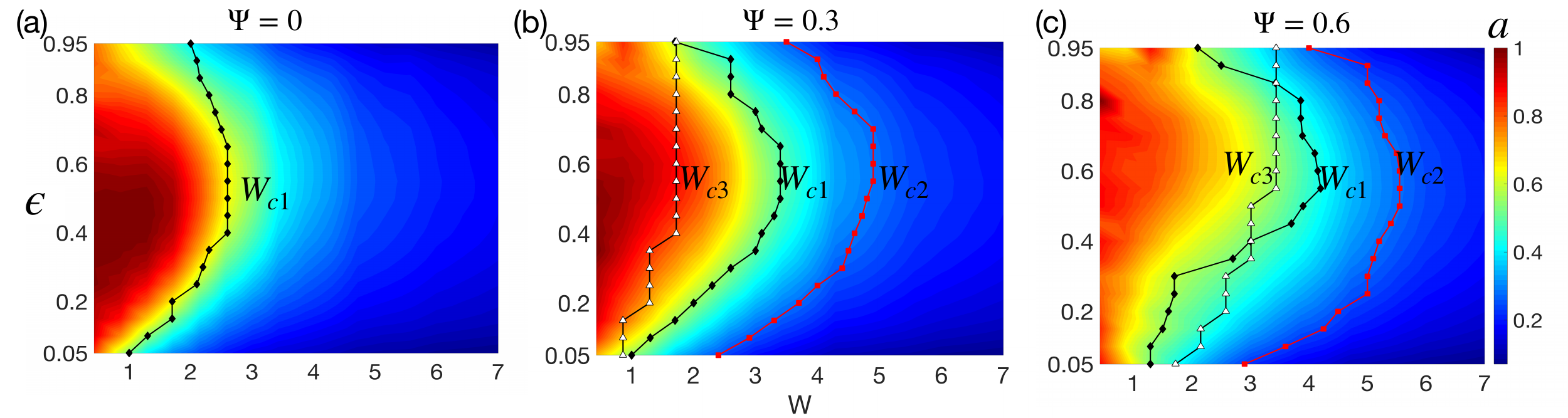}}
	\caption{\textbf{Phase diagrams:}  The phase diagrams for (a) the Hermitian system $\Psi=0$, and the non-Hermitian systems (b) $\Psi=0.3$ and (c) $\Psi=0.6$, as a function of disorder ($W$) and energy density $(\epsilon)$. The color indicates the slope $a$ of the participation entropy with the system size (see main text). The slope exhibits transition between ergodic to non-ergodic eigenstates, going from $a\approx 1$ deep in the ergodic phase to $a\ll 1$ deep in the non-ergodic (MBL) phase. The transition shifts to higher disorder with increasing $\Psi$, indicating breakdown of MBL states via current drive. The phase boundaries -- (i) $W_{c1}$, Hilbert space delocalization-localization and entanglement transitions, (ii) $W_{c2}$, time-reversal breaking transition, and (iii) $W_{c3}$, current transition, are shown. Eigenstates have volume-law entanglement and are delocalized for $W<W_{c1}$, whereas, for $W>W_{c1}$, states have area-law entanglement and are localized.  A finite fraction of eigenstates breaks $\mc{T}$-reversal for $W<W_{c2}$. The region, $W_{c1}<W<W_{c2}$, have area-law localized states that break $\mc{T}$-reversal. There is a separate transition at $W_{c3}$ in terms of system-size scaling of current (see main text).}
	\label{fig:PhaseDiagram} 
\end{figure*}   
We discuss our results on the phases and phase diagram for the model of Eq.\eqref{eq.SpinModel} in this section. To this end, we obtain the eigenvalues and the (left and right) eigenvectors of the non-Hermitian Hamiltonian [Eq.\eqref{eq.SpinModel}] using numerical exact diagonalization for system sizes $L=10-16$ and sample over many disorder realizations (10000 for $L=10$, 6000 for $L=12$, 1440 for $L=14$, and 200 for $L=16$). The Hamiltonian [Eq.\eqref{eq.SpinModel}] and the dynamics [Eq.\eqref{eq.DynEq}] conserves $S_z^{tot}=\sum_i S_i^z$. Hence we work in the $S_z^{tot}=0$ subspace. We expect the results to be similar for the other $S_z^{tot}$ sectors, however the effect of interaction is strongest for $S_z^{tot}=0$ subspace, which corresponds to half filling in the fermion sector. We take the interaction $J_z=J=1$ and vary $W$ and $\Psi$.

As in the studies of MBL systems \cite{Oganesyan2007,Pal2010,Altman2015,Nandkishore2015,Abanin2018,Luitz2015,Mace2019}, here we characterize the phases as a function of $W$ and $\Psi$ in terms of (a) properties of eigenstates and (b) from time evolution starting from a generic initial pure state.  Moreover, to draw the eigenstate phase diagram we use the quantity, $\epsilon=(\mc{E}-\mc{E}_0)/(\mc{E}_M-\mc{E}_0)$, defined from the real part of the eigenvalues; $\mc{E}_M$ and $\mc{E}_0$ are the maximum and the minimum values of $\mc{E}$, respectively. For brevity, we refer to $\epsilon$ as \emph{energy density}. The quantity $\epsilon$ is used as an index for the eigenstates. In the Hermitian limit $\Psi=0$, since energy is conserved under time evolution, the expectation value of energy density $\epsilon$ of the initial state can be used to characterize dynamics and long-time steady state ensemble. For example, in the ergodic phase of the Hermitian model, within eigenstate thermalization hypothesis (ETH)\cite{Deutsch1991,Srednicki1994,Srednicki1999}, the long-time steady state can be related to a thermal Gibbs ensemble at a temperature $T$ with the thermal expectation value of energy density $\epsilon(T)=\epsilon$. Similarly, this relation can be used to define a temperature for an eigenstate using its energy density. As a result, an eigenstate phase transition at the many-body mobility edge ($\epsilon_c$) from ergodic to non-ergodic or MBL phase can be defined as a function of $\epsilon$ \cite{Basko2006,Gornyi2005,Oganesyan2007,Pal2010,Altman2015,Nandkishore2015,Abanin2018,Luitz2015,Mace2019}. In the non-Hermitian case the energy density $\epsilon$ is not conserved under the dynamics [Eq.\eqref{eq.DynEq}] and for evolution from a generic initial pure state the energy of the long-time steady state is uniquely determined by the energy density of the eigenstate, $|s_R\rangle$, as can be easily seen from Eq.\eqref{eq:NESS}. Hence $\epsilon$ cannot be related with a temperature. However, $\epsilon$ can still be used to denote the spectrum of states and eigenstate transitions, e.g.  $\mc{T}$-reversal breaking and localization-delocalization transition or a many-body mobility edge, analogous to the single-particle mobility edge in the non-interacting Hatano-Nelson model \cite{Hatano1996,Hatano1997,Hatano1998}. Furthermore, states other than $|s_R\rangle$ are accessed during the short and intermediate-time dynamics and they govern the approach towards the steady state as we discuss later. 

Following the above considerations, we construct the eigenstate phase diagram in the $W-\epsilon$ plane for a few values of $\Psi$ based on -- (a) participation entropy \cite{Luitz2015,Mace2019}, which quantifies the extent of delocalization of an eigenstate in the Hilbert space for a chosen basis, as discussed in the next section, (b) entanglement entropy, and (c) fraction of eigenstate with non-zero imaginary part of eigenvalue. The last one serves as an order parameter for $\mc{T}$-reversal symmetry breaking in the eigenstates. In addition, we further quantify the $\mc{T}$-reversal breaking via scaling of current carried in the eigenstate with system size. As mentioned in the introduction, we find eigenstate transitions in terms of all these quantities. We assume the transitions to be continuous as in the numerical studies of MBL transition for the Hermitian case and corroborate the assumption via finite-size scaling collapse of the data of various quantities. However, we note that, even in the Hermitian case, the exact nature of MBL transition is currently debated \cite{Vosk2015,Potter2015,Dumitrescu2017,Thiery2018,Goremykina2019,Roy2019,Roy2019a}, and cannot be ascertained within the modest system sizes accessed via  ED studies.  


To characterize the phases based on time evolution, we look into the long-time steady state and the time evolutions of entanglement entropy and N\'{e}el order parameter, which serves as the MBL order parameter \cite{Schreiber2015}, starting from an initially un-entangled N\'{e}el state of staggered arrangement of spins. We describe our results in detail in the next sections.  

\begin{figure}[!htb]
	\centering{
		\includegraphics[width=\linewidth,height=55mm]{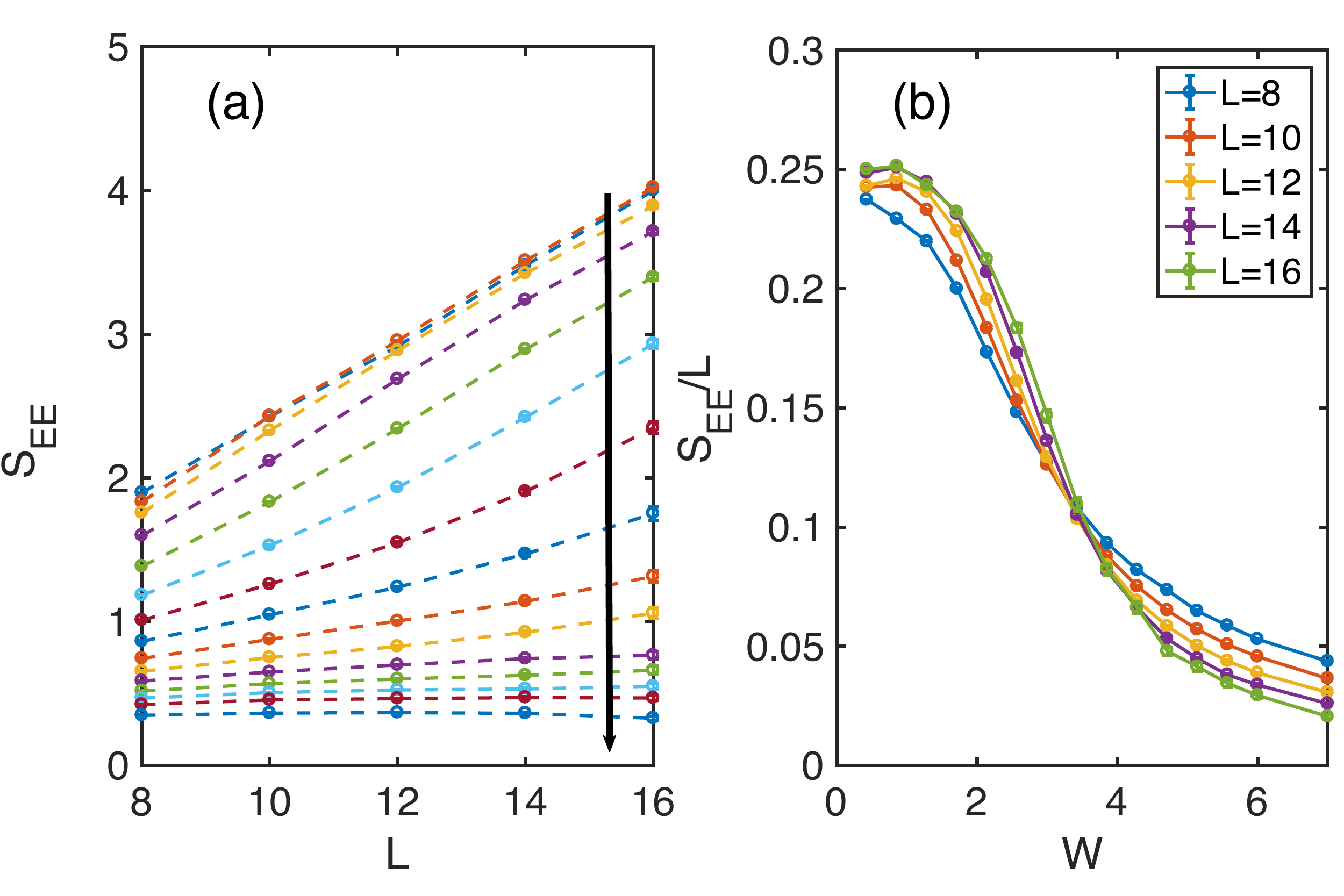}}	
	\caption{\textbf{Volume-law to area-law transition for $\Psi=0.3$:} (a) $S_{EE}$ as function of $L$ for different disorder strengths at $\epsilon=0.5$; the arrow denotes the direction of increasing disorder ($W$), from 0.43 to 7. (b) $S_{EE}/L$ is plotted against $W$ for different $L$. The crossing of the curves for different $L$ at $W\simeq 3.6$ suggests the existence of a scale invariant critical point.}
	\label{fig:a2v} 
\end{figure}
\subsection{Ergodic to non-ergodic eigenstate transition in the Hilbert space}  

To characterize the eigenstates of the non-Hermitian Hamiltonian [Eq.\eqref{eq.SpinModel}], we first obtain a phase diagram in terms of participation entropy, a diagnostic of ergodicity of the eigenstates. It's defined as $S_P(|n_R\rangle)=-\sum_\alpha p_\alpha^{(n)} \ln{p_\alpha^{(n)}}$ for the eigenstates in the basis of the spin configurations, $|\alpha\rangle=|S_1^z,S_2^z,\dots,S_N^z\rangle$  \cite{Luitz2014,Luitz2015} with $p_\alpha^{(n)}=|\langle \alpha|n_R\rangle|^2$.


In the delocalized phase, eigenstates are ergodic and has support over a finite fraction of sites in the Hilbert space, i.e. $\langle \alpha|n_R\rangle\sim1/\sqrt{D_H}$ as required by normalization, and hence $S_P\simeq \ln{D_H}$, where $D_H$ is the dimension of the Hilbert space in the $S_z^{tot}=0$ subspace. On the other hand, in the MBL phase, the eigenstates are expected to exhibit a fractal character, i.e. a delocalized but non-ergodic behavior, with support over exponentially large number but vanishing fraction of sites in the Hilbert space, i.e. $\langle \alpha|n_R\rangle\sim 1/D_H^{a/2}$ or $S_P\simeq a\ln{D_H}$ with $a<1$ \cite{Mace2019,Logan2019}. We obtain the disorder averaged $S_P(\epsilon)$ as a function $W$ and $\epsilon$ and plot the slope of $S_P$ with $L$ ($\sim \ln{D_H}$) in Fig.\ref{fig:PhaseDiagram} (a-c) for $\Psi=0,0.3$ and 0.6. To define $S_P(\epsilon)$, and other quantities discussed in the next sections, as a function of $\epsilon$, we have binned $\epsilon$ into intervals containing 100 or more eigenstates for a given disorder realization. Indeed, consistent with the Hermitian MBL case \cite{Luitz2015,Mace2019} in Fig.\ref{fig:PhaseDiagram} (a), we find a transition in the Hilbert space in terms of eigenstate participation for $\Psi=0.3,0.6$ Fig.\ref{fig:PhaseDiagram} (b),(c). The transition shifts to higher disorder for increasing $\Psi$. Hence, a non-zero current drive causes a \emph{breakdown} of the MBL states of the parent Hermitian model over certain regime of disorder.

\begin{figure}[!htb]
	\centering{
		\includegraphics[width=\linewidth,height=110mm]{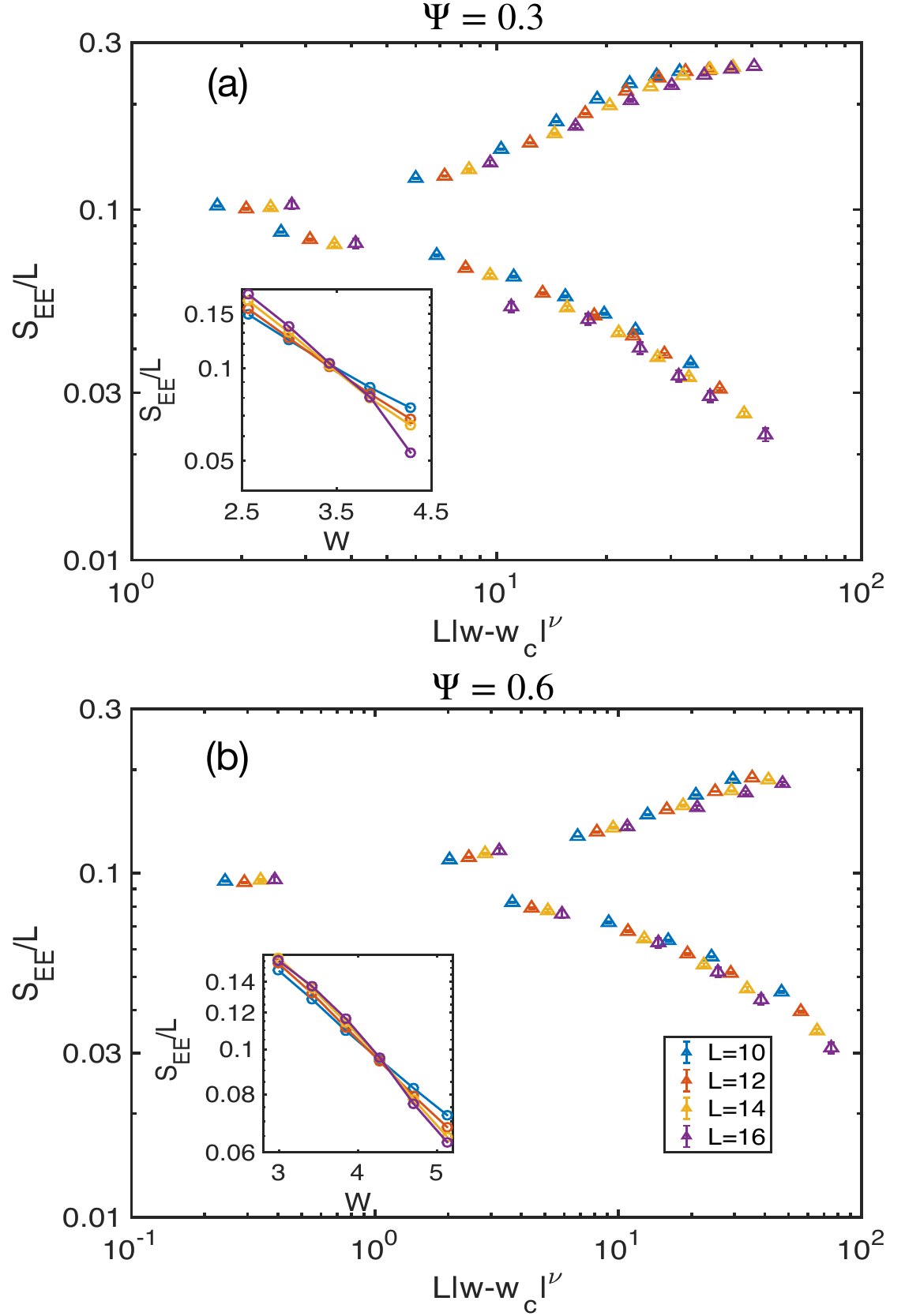}}
	\caption{\textbf{Finite-size scaling of entanglement entropy:} The scaling collapse of $S_{EE}/L$ is shown at $\epsilon=0.5$ for the non-Hermitian model for (a) $\Psi=0.3$ and (b) $\Psi=0.6$. In both the cases, the data for different $L$ collapses reasonably well into two universal curves, one below the transition ($W<W_{c1}$) and the other above ($W>W_{c1}$). The critical disorder $W_{c1}$ in each case has been extracted from the system size crossing of $S_{EE}/L$ as shown in the insets. The values of critical disorder and scaling exponents extracted for different $\Psi$ are, 
(a) $\nu_1\approx 1,W_{c1}=3.6\pm0.1$ for $\Psi=0.3$, and (b) $\nu_1\approx 1.5,W_{c1}=4.2\pm0.1$ for $\Psi=0.6$.} \label{fig:ScalingEntanglement} 
\end{figure}   

\begin{figure}[!htb]
	\centering{
		\includegraphics[width=\linewidth,height=120mm]{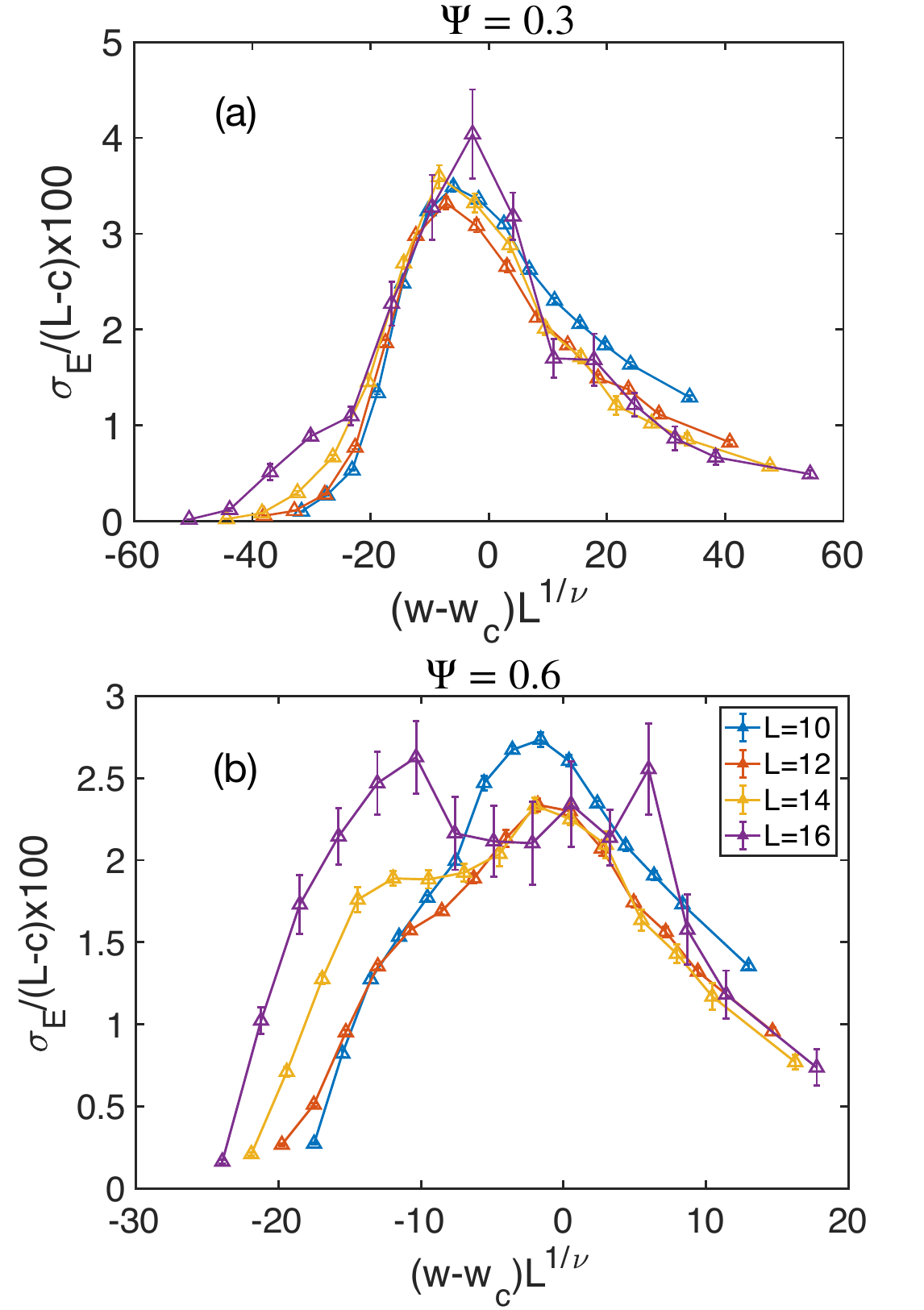}}
	\caption{\textbf{Finite-size scaling of variance of entanglement entropy:} System size scaling of $\sigma_E$ is shown here at $\epsilon=0.5$ for $\Psi=0.3$ and $\Psi=0.6$ in panel (a) and (b), respectively. The critical disorder $W_{c1}$ in each case has been fixed at those from Fig.\ref{fig:ScalingEntanglement}. The scaling of the peak height and position indicate the existence of an entanglement transition at $W_{c1}$ in tune with the results in Fig.\ref{fig:ScalingEntanglement}.}
	\label{fig:Scalingvariance} 
\end{figure}  
\subsection{Entanglement transition} 
 We obtain the von Neumann entanglement entropy $S_{EE}$ for each the eigenstates, i.e. $S_{EE}(|n_R\rangle)=-\mr{Tr}(\rho_A\ln\rho_A)$. Here $\rho_A=\mr{Tr}_B|n_R\rangle\langle n_R|$ is the reduced density matrix of the subsystem $A$, for the real-space bipartition of the system into left half, $A$, and right half, $B$. As already discussed in the introductory section, the MBL transition is defined from volume-law to area-law transition \cite{Luitz2015} in the Hermitian case, i.e. $\Psi=0$. A similar transition is  observed for $\Psi=0.3$ with increasing disorder as shown in Fig.\ref{fig:a2v}(a), where disorder averaged $S_{EE}$ is plotted as a function of length $L$ at $\epsilon=0.5$, at the middle of the spectrum, for a range of disorder strength. As can be seen, for large values of disorder $S_{EE}$ tends to a constant value at larger $L$, whereas $S_{EE}$ increases linearly with $L$ for weaker disorder strength. The presence of a transition is evident from a clear crossing of $S_{EE}/L$ vs.~$W$ curves for different $L$ in Fig.\ref{fig:a2v}(b). The dependence of $S_{EE}$ on $L$ is consistent with a volume-law scaling for $W<W_{c1}$, and an area-law scaling for $W>W_{c1}$. This crossing point is used to find out the critical disorder $W_{c1}\approx 3.6$, implying an entanglement transition, similar to the Hermitian case \cite{Luitz2015}.
 
 The transition is further corroborated by a reasonably good data collapse in Fig.\ref{fig:ScalingEntanglement}(a),(b) for $\Psi=0.3,~0.6$, obtained using the finite-size scaling ansatz $S_{EE}(L)=Lg(L^{1/\nu_1}(W-W_{c1}))$ \cite{Luitz2015}, where $g(x)$ is the scaling function. The finite-size sclaing form assumes a volume-law scaling at the critical point, as in the case of Hermitian model \cite{Luitz2015}. We find that the data for $S_{EE}/L$ at $\epsilon=0.5$ and $\Psi=0.3$ [Fig.\ref{fig:ScalingEntanglement}(a)] for various $W$ and $L$ can be collapsed quite well into two universal curves for $W<W_{c1}$ and $W>W_{c1}$ with a critical exponent $\nu_1\simeq 1$ and $W_{c1}=3.6\pm 0.1$. The two curves, one below and another above the transition, result from the sign of $W-W_{c1}$ in the argument of the scaling function $g(x)$. We have done similar analysis for the entire spectrum of $\epsilon$. The crossing of the curves and the data collapse are most prominent near the middle of the spectrum. The finite-size data collapse of $S_{EE}/L$ at $\epsilon=0.5$ for $\Psi=0.6$ is shown in Fig.\ref{fig:ScalingEntanglement}(b). For this larger value of $\Psi$, we find $W_{c1}=4.2\pm 0.1$, and $\nu_1=1.5$, i.e. the extracted critical exponent changes with $\Psi$. 
 
 Here it is important to note that, as in the exact diagonalization studies of Hermitian MBL systems \cite{Luitz2015}, for $S_{EE}(L)$ and the other quantities discussed in the next sections, neither the scaling function $g(x)$ nor the critical exponents can be extracted very accurately from finite-size scaling limited to such small systems, and with only a few system sizes accessible via ED. However, the scaling function $g(x)$, as such, is not required to obtain the data collapse. The collapse is obtained by taking all the data as function of $W$ and $L$, e.g. from Fig.\ref{fig:ScalingEntanglement}(b) (inset), and plotting as a function of the $L|W-W_{c1}|^{\nu_1}$ (or $L^{1/\nu_1}|W-W_{c1}|$) with appropriate choice of $W_c$ and $\nu$ that generate a good scaling collapse. Additionally, we have also used a polynomial for $g(x)$ and fitted all the data as a function of $L^{1/\nu_1}(W-W_{c1})$ with the polynomial. This gives results consistent with that obtained using the simpler data collapse mentioned above. We do not perform a detailed error analysis for the scaling collapse. The main purpose of the finite-size scaling here is to bring out the existence of the transition, rather than estimating the critical exponents. The errorbars in $W_{c1}$ is estimated from the range of values of $W_{c1}$ for which the data collapse is reasonable while keeping the critical exponent $\nu_1$ fixed. The finite-size scaling analysis for the other quantities discussed below are also performed in a similar manner.  

For the the phase boundaries $W_{c1}$ in the $W-\epsilon$ plane for $\Psi=0,0.3,0.6$ in Fig.\ref{fig:PhaseDiagram} we use the standard deviation, $\sigma_E$, of the entanglement entropy over disorder realizations. At the transition, $\sigma_E$ is expected to show a peak that diverges with $L$ \cite{Kjall2014}. In Fig.\ref{fig:PhaseDiagram}, we plot $W_{c1}$ from the peak of $\sigma_E$ for system size $L=16$. The phase boundary is consistent with that obtained from the participation entropy for ergodic to non-ergodic transition. In Figs.\ref{fig:Scalingvariance}(a) and (b), we obtain scaling collapses of $\sigma_E/(L-c)$ vs.~$W$ for $\Psi=0.3$ and $\Psi=0.6$ at $\epsilon=0.5$ with $c$ as a fitting parameter \cite{Luitz2015}. Here the values of $\nu_1$ and $W_{c1}$ are fixed from the finite-size scaling of $S_{EE}$. Surprisingly we see an indication of two peaks in $\sigma_E$ for $L = 16$ [Fig.\ref{fig:Scalingvariance}(b)]. The reason behind the two-peak structure is not clear at present and will be studied in a future work.

We note that the average $S_{EE}$ and $\sigma_E$ shown in Fig.\ref{fig:ScalingEntanglement} and Fig.\ref{fig:Scalingvariance} are obtained by averaging over the eigenstates with real eigenvalues. The averaging over the eigenstates with complex eigenvalues also gives similar results.


\subsection{$\mc{T}$-reversal breaking:} 
\begin{figure}[!htb]
\begin{center}
\includegraphics[width=\linewidth,height=115mm]{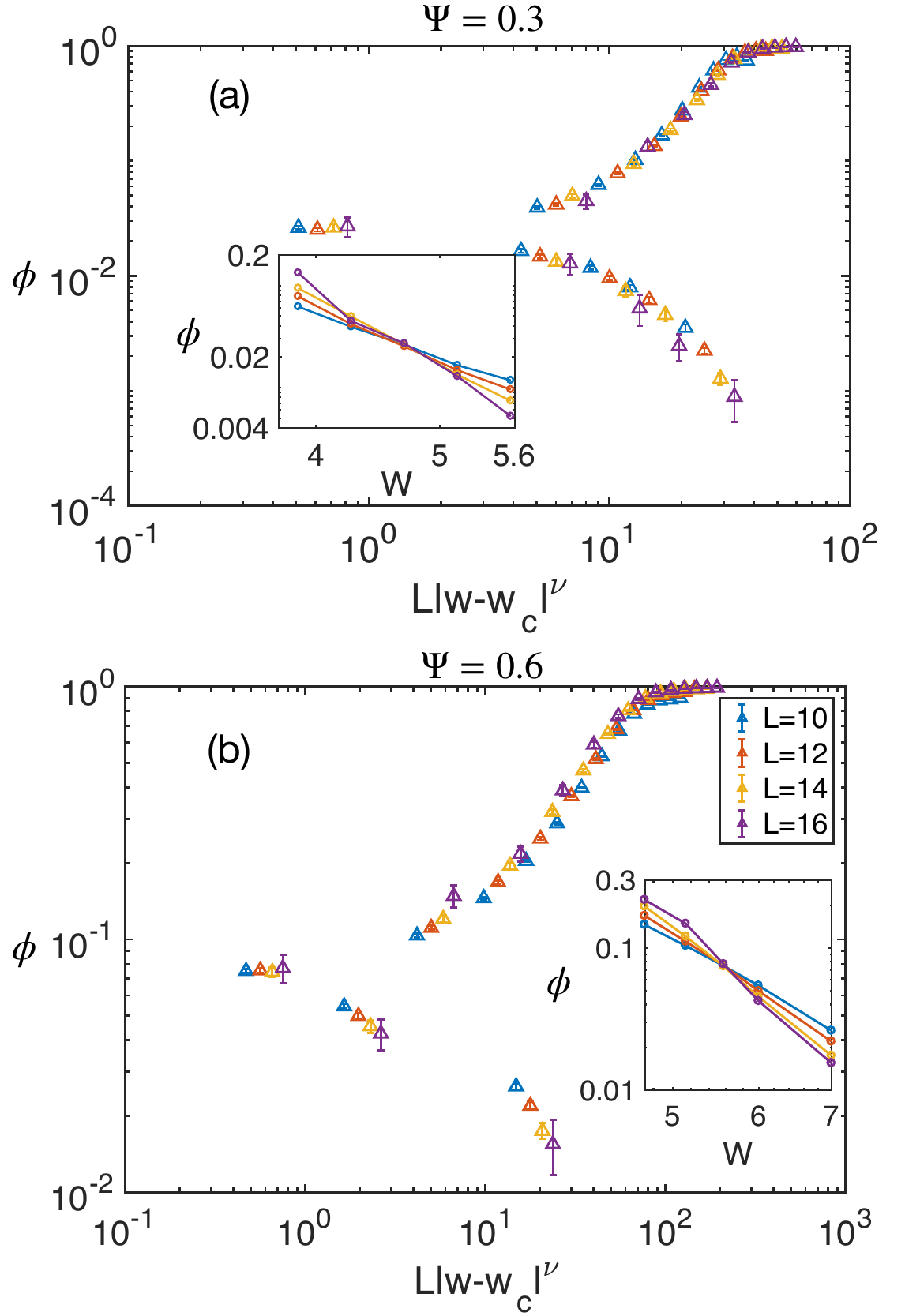}
\end{center}
	\caption{\textbf{Time-reversal symmetry breaking transition:} System size scaling of fraction of imaginary eigenvalues ($\phi$) at $\epsilon=0.5$ for (a) $\Psi=0.3$, and (b) $0.6$. The critical disorder $W_{c2}$ is extracted from the system size crossing of $\phi$ as shown in the insets. The values of critical disorder and scaling exponents extracted for different $\Psi$ are, (a) $\nu_2\approx 0.9,W_{c2}=4.75\pm0.1$ for $\Psi=0.3$, (b) $\nu_2\approx 1.5,W_{c2}=5.6\pm0.1$ for $\Psi=0.6$.  The data for different $L$ collapses into two universal curves, one below the transition ($W<W_{c2}$) and the other above ($W>W_{c2}$). This, and the fact that $W_{c2}>W_{c1}$, indicate the presence of $\mc{T}$-reversal breaking transition distinct from the entanglement transition in Fig.\ref{fig:ScalingEntanglement}.}
	\label{fig:TBreaking} 
\end{figure} 

We now address one of the main questions raised in the introductory section, i.e., whether the entanglement or the localization transition coincides with the $\mc{T}$-reversal breaking, as in the non-interacting Hatano-Nelson model \cite{Hatano1996,Hatano1997,Hatano1998}. We define a $\mc{T}$-reversal \emph{order parameter}, $\phi(\epsilon)$, the fraction of imaginary eigenvalues at $\epsilon$. Numerically an eigenvalue is defined to have a non-zero imaginary part  by setting an ad-hoc small cutoff. However, we have checked that our results are insensitive to the choice of cutoff. We find a clear crossing of the $\phi$ vs.~$W$ curves for different $L$, as shown in Fig.\ref{fig:TBreaking}(a),(b) (insets) for $\Psi=0.3$ and $\Psi=0.6$ at $\epsilon=0.5$. As evident, the crossing point is at $W=W_{c2}\approx 4.75$ for $\Psi=0.3$, clearly larger than $W_{c1}$ for the entanglement transition in Fig.\ref{fig:ScalingEntanglement}(a). A good scaling collapse can again be obtained with an exponent $\nu_2\simeq 0.9$  and $W_{c2}=4.75\pm 0.1$ [Fig.\ref{fig:TBreaking}(a)]. The collapse of the data for $\phi$ vs.~$W$ could not be obtained with $W_{c2}=W_{c1}$. This establishes the fact that $\mc{T}$-reversal breaking is distinct from the entanglement transition and occurs within the area-law phase. We find similar results for $\Psi=0.6$ [Fig.\ref{fig:ScalingEntanglement}(b) and Fig.\ref{fig:TBreaking}(b)], namely $W_{c2}=5.6\neq W_{c1}$. In this case, the critical exponent $\nu_2=1.5$. From the crossing points of $\phi(\epsilon)$ vs.~$W$ curves we obtain the phase boundary $W_{c2}$ for the $\mc{T}$-reversal breaking in Figs.\ref{fig:PhaseDiagram}. 

\begin{figure}[!htb]
	\centering{
		\includegraphics[width=\linewidth,height=115mm]{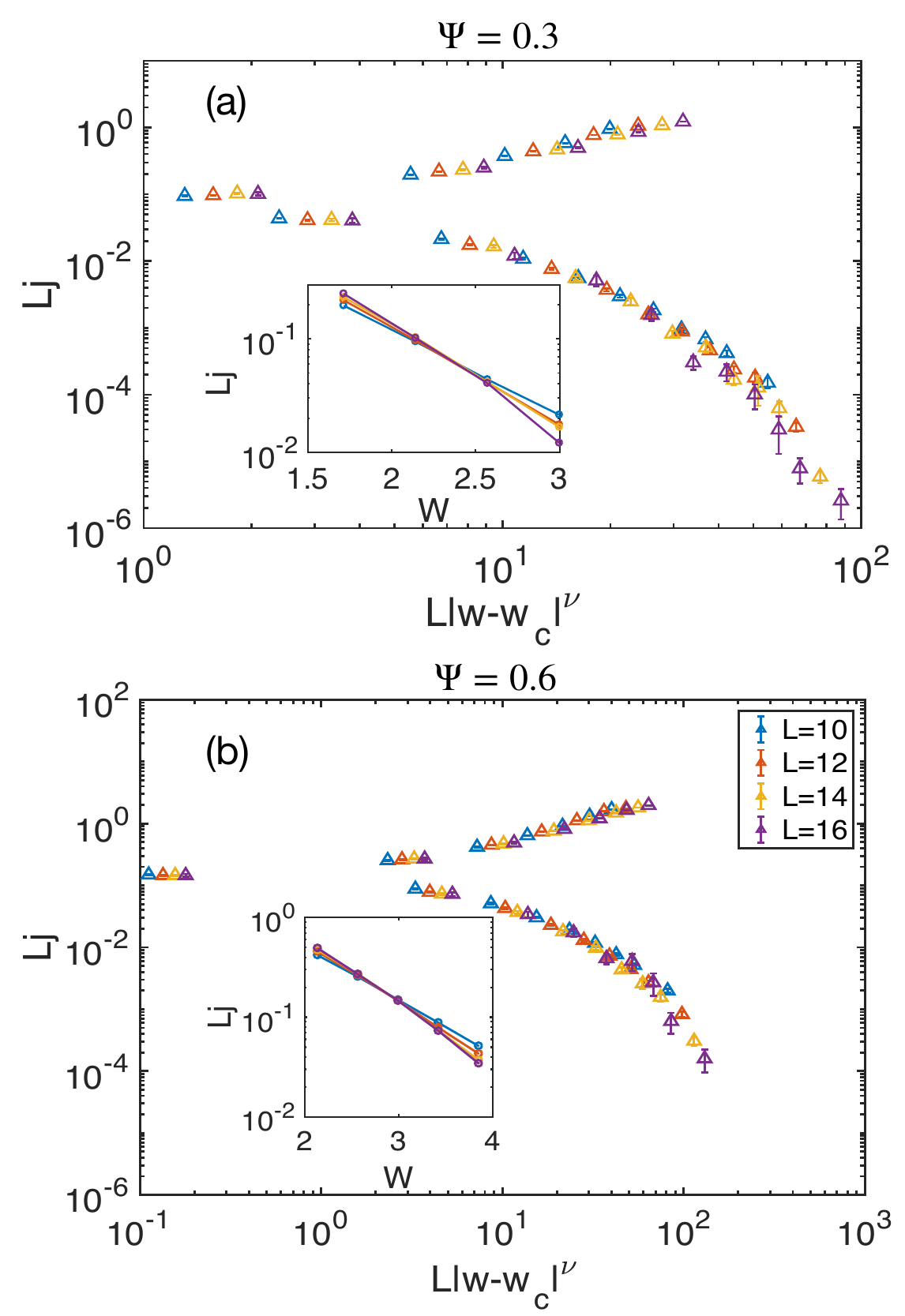}}
	\caption{\textbf{Finite-size scaling of average eigenstate current:} System size scaling of average current($j$) is shown here for $\epsilon=0.5$ and $\Psi=0.3,0.6$ in panels (a) and (b), respectively. The critical disorder $W_{c3}$ in each case has been extracted from the system size crossing of $Lj$ as shown in the insets. The values of critical disorder and scaling exponents extracted for different $\Psi$ are, (a) $\nu_2\approx 1,W_{c2}=2.3\pm0.1$ for $\Psi=0.3$, (b) $\nu_3\approx 1.5,W_{c3}=2.95\pm0.1$ for $\Psi=0.6$. The data for different $L$ collapses into two universal curves, one below the transition ($W<W_{c3}$) and the other above ($W>W_{c3}$) indicating a transition in terms of system-size scaling of current. The transition at $W_{c3}\neq W_{c1},W_{c2}$ is distinct from either the entanglement transition [Fig.\ref{fig:ScalingEntanglement}] or the $\mc{T}$-reversal breaking transition [Fig.\ref{fig:TBreaking}]. }
	\label{fig:ScalingCurrent}	
\end{figure} 

To further characterize the $\mc{T}$-reversal breaking eigenstates, we compute the current $j(\epsilon)$, obtained by averaging over the magnitude of the currents, $|j_n|$, carried by the eigenstates with imaginary eigenvalues, and disorder realizations. Again, we find crossings, $W=W_{c3}\approx 2.3$ for $\Psi=0.3$ and $W=W_{c3}\approx 3.0$ for $\Psi=0.6$, in the $\mc{J}=Lj$ vs.~$W$ plots for different $L$ [Figs.\ref{fig:ScalingCurrent}(a),(b)] at $\epsilon=0.5$. The transition, which we refer to as \emph{current transition} for brevity, is seemingly distinct from both entanglement and $\mc{T}$-reversal breaking transitions. We can obtain a scaling collapse for $j$ with $\nu_3=1$  and $W_{c3}=2.3$ for $\Psi=0.3$, and $\nu_3=1.5$  and $W_{c3}=3.0$ for $\Psi=0.6$, as shown in Figs.\ref{fig:ScalingCurrent}(a),(b)(inset). We obtain a phase boundary for the current transition at other values of $\epsilon$ from the crossing points, as shown in Figs.\ref{fig:PhaseDiagram}(b),(c). 

 The scaling of $j$ with $L$ for $W\ll W_{c3}$ is consistent with $j$ approaching a constant for $L\to \infty$ (not shown). The scale-invariant crossing point indicates a diffusive scaling of the current at the transition, namely $j\sim 1/L$. In fact, for $W_{c3}<W\aplt W_{c2}$, the scaling of $j$ with $L$ could be consistent with $j\sim 1/L^\gamma$ with $\gamma \apgt 1$, and we expect $j\sim e^{-L/\zeta}$ for $W\gg W_{c2}$, deep inside the $\mc{T}$-reversal unbroken localized phase; $\zeta$ is the characteristic localization length. However, this is hard to verify from the exact diagonalization numerics limited to such small system sizes. As discussed later [see Fig.\ref{fig:jinf}], the current, $j_\infty$, carried by the long-time NESS also exhibits similar transition. 
 For $W>W_{c1}$, in the area-law phase, the scaling of the current and the transition at $W_{c2}$ could be studied by a matrix-product operator (MPO) based implementation \cite{Prosen2009,Znidaric2014} of the dynamical evolution in Eq.\eqref{eq.DynEq}. It would be interesting to establish the existence of such current-carrying pure NESS with area-law entanglement in an interacting system, as discussed in ref.\onlinecite{Gullans2018}. 

   \begin{figure}[!htb]
 	\centering{
 		\includegraphics[width=\linewidth,height=110mm]{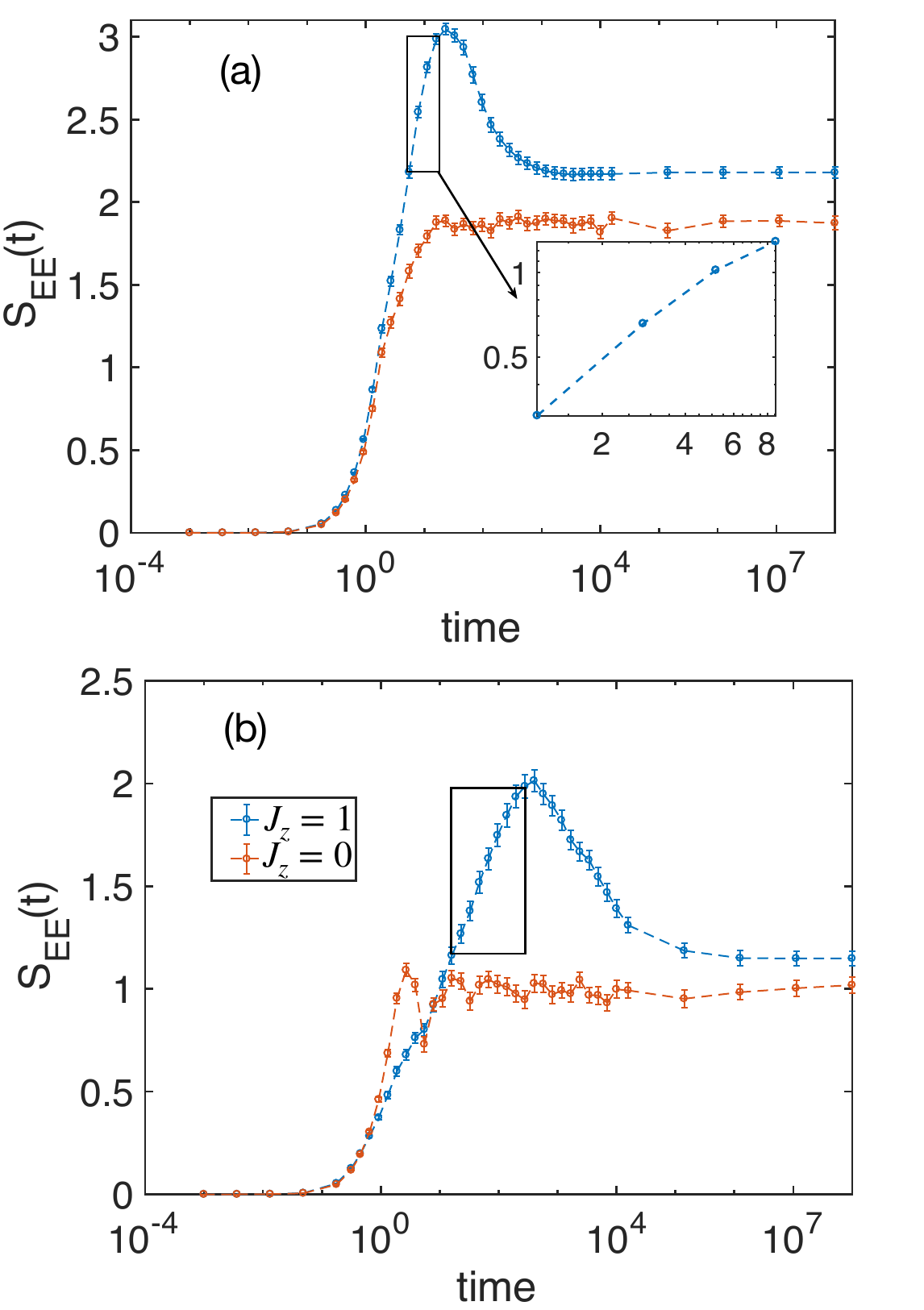}}
 	\caption{\textbf{Time evolution of entanglement entropy:} (a), (b) Semilog plots of entanglement entropy $S_{EE}$ vs. $t$ for $W=2.15$, in the $\mc{T}$-reversal broken volume-law phase, and $W=4.3$, within the $\mc{T}$-reversal broken area-law phase, respectively, starting with the Ne’el state, for the interacting, $J_z=1$ ($S_{EE}^{J_z=1}$), and the non-interacting, $J_z=0$ ($S_{EE}^{J_z=0}$), cases. Inset in (a) shows log-log plot of $S_{EE}^{J_z=1}(t)-S_{EE}^{J_z=0}(t\rightarrow \infty)$ vs. $t$. The box in (b) indicates the initial logarithmic growth in the localized phases, with $\mathcal{T}$-reversal breaking. The entanglement entropy peaks at a finite time and then eventually decays to a finite value corresponding to the NESS (see main text). 
 	}
 	\label{fig:Entanglementgrowth}
\end{figure}  
 
 The phase boundaries in the $W-\epsilon$ plane for all the transitions, ergodic to non-ergodic, entanglement, time-reversal breaking and current transition, and their evolutions with $\Psi$ are summarized in Fig.\ref{fig:PhaseDiagram}(b),(c). For comparison, we also show the ergodic to non-ergodic and entanglement transition for $\Psi=0$ in Fig.\ref{fig:PhaseDiagram}(a). The three phase boundaries $W_{c1},~W_{c2}$ and $W_{c3}$ naturally shift to higher disorder with increasing $\Psi$. As already mentioned, we obtain the phase boundary, $W_{c1}(\epsilon)$, from the peak of the standard deviation of entanglement entropy $\sigma_E$ [Fig.\ref{fig:Scalingvariance}. Similar, albeit slightly higher, values for $W_{c1}$ are obtained from the crossing of $S_{EE}/L$ vs.~$W$ curves [Fig.\ref{fig:ScalingEntanglement}]. However, we do not use the crossing point to plot phase boundary as a function of $\epsilon$, since clear crossings can only be detected over middle one-third and one-half of the spectra for $\Psi=0.3$ and $\Psi=0.6$, respectively. We note that the clear crossing point for $S_{EE}/L$ vs.~$W$ curves could be obtained for the Hermitian case ($\Psi=0$) almost over the entire spectrum, except at the very edges.

 \subsection{Growth and decay of entanglement entropy and the approach to NESS:} \label{subsec:EntanglementDynamics}
 In this section we study the time evolution under the non-Hermitian evolution [Eq.\eqref{eq:purestate}] starting with a simple un-entangled initial state. We choose the N\'{e}el state $|\psi_0\rangle=|\uparrow\downarrow\uparrow\downarrow\dots\rangle$ for this purpose. We compute the entanglement entropy, $S_{EE}(t)$, as function of time from the time-evolved state $|\psi(t)\rangle$ [Eq.\eqref{eq:purestate}]. The results for $S_{EE}(t)$ are shown in Figs.\ref{fig:Entanglementgrowth}(a),(b) for two values of disorder strength, (a) $W=2.15$, in the volume-law phase, and (b) $W=4.3$, the $\mc{T}$-reversal broken area-law phase. The result for higher disorder, in the $\mc{T}$-reversal unbroken area-law phase, is similar to that for (b). In each of the cases, to bring out the crucial effect of interaction, we compare $S_{EE}(t)$ with that obtained for the non-interacting Hatano-Nelson model ($J_z=0$). The prominent features of $S_{EE}(t)$ are, (i) an initial growth, (ii) a broad peak followed by a decay or relaxation, and (iii) an eventual approach to a steady state value corresponding to the NESS. In the volume-law phase [Fig.\ref{fig:Entanglementgrowth}(a)(inset)], $S_{EE}(t)$ grows linearly with time, whereas, $S_{EE}(t)\propto \ln{t}$ initially in the area-law phase [Fig.\ref{fig:Entanglementgrowth}(b)], as in the Hermitian MBL case \cite{Bardarson2012}. 
 
 Using Eq.\eqref{eq:purestate}, the growth and subsequent decay of entanglement entropy can be understood from the density matrix, $\rho(t)=|\psi(t)\rangle\langle \psi(t)|=e^{-\mc{L}t}\rho_0=\sum_{n,m} e^{-(\beta_{nm}+\ci\delta_{nm})t}C^\psi_{nm}(t)|n_R\rangle \langle m_R|$, where the coefficient $C^\psi_{nm}(t)$ is obtained from Eq.\eqref{eq:purestate}.
Here $-(\beta_{nm}+\ci\delta_{nm})$ could be thought of as the eigenvalues of the Liouvillian operator $\mc{L}$; $\beta_{nm}=2\Lambda_s-\Lambda_n-\Lambda_m\geq 0$ and $\delta_{nm}=\mc{E}_n-\mc{E}_m$. The real part, $-\beta_{nm}$, of eigenvalue of $\mc{L}$ leads to relaxation and $\beta_{nm}=0$ corresponds to the long-time steady state. For a weak strength ($\Psi\ll 1$) of the non-Hermitian term, and for $t\ll e^{L}$ \footnote{Since the many-body spectra for the real part of the eigenvalues are exponentially dense, the actual typical level spacing $\tilde{\delta}_{typ}\sim e^{-L}$. However this does not contribute to the time evolution in Figs.\ref{fig:Entanglementgrowth}(a), (b) for $t\ll e^L$}, $\delta_{typ}\gg \beta_{typ}$, and the initial growth of the entanglement entropy appears over a time window, $\delta_{typ}^{-1}\aplt t \aplt \beta_{typ}^{-1}$, due to the dephasing from the exponential factor $e^{\ci\delta_{nm}t}$ in $\rho(t)$. In this time window, the factor $e^{-\beta_{nm}t}\approx 1$. Here $\delta_{typ}$ and $\beta_{typ}$ are the typical values of $\delta_{nm}$ and $\beta_{nm}$, respectively, that contributes to $\rho(t)$ for $t\ll e^{L}$. 

 \begin{figure*}[!htb]
 	\centering{
 		\includegraphics[width=\linewidth,height=45mm]{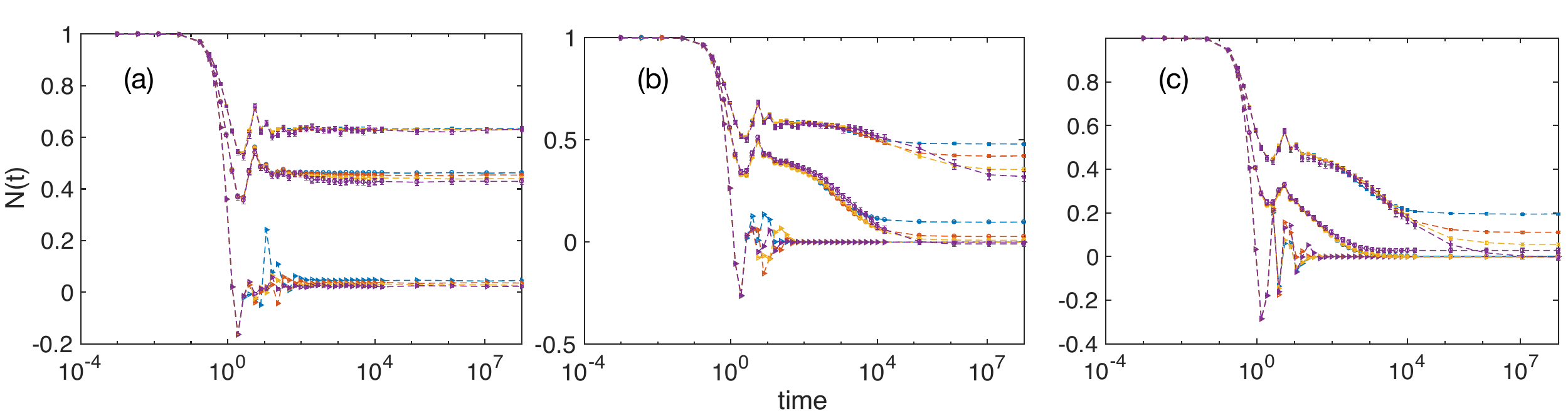}}
 	\caption{\textbf{Time evolution of N\'{e}el order parameter:} The N\'{e}el order parameter $N(t)$ is shown for (a), the Hermitian case, $\Psi=0$, and the non-Hermitian cases, (b) $\Psi=0.3$ and (c) $\Psi=0.6$ as a function of $W$ and $L$. The triangles, circles and squares are for $W = 0.43, 4.3, 7$, respectively, with $L = 10$ (blue), $L = 12$ (orange), $L = 14$ (yellow) and $L = 16$ (purple). The errorbars are smaller than the symbols. In the Hermitian MBL phase $N_\infty=N(t\rightarrow \infty)$ remains finite in the limit $L\rightarrow \infty$, whereas $N_\infty$ decays with increasing $L$ in the non-Hermitian MBL phase, indicating the loss of memory of initial state in the thermodynamic limit. A new dynamical regime emerges in the non-Hermitian MBL phase where the $N(t)$ plateaus over a long intermediate time window retaining the initial memory.}
 	\label{fig:NeelOrderParameter} 
 \end{figure*} 

This initial dephasing mechanism is similar to the one that leads to $\ln{t}$-growth of entanglement entropy in the MBL phase \cite{Serbyn2013}. For the interacting system, left to itself, the dephasing would typically lead to a diagonal ensemble and long-time state with volume-law entanglement, as it does for the MBL phase \cite{Bardarson2012}. 
Nevertheless, for the non-Hermitian case, the relaxation due to $\beta_{nm}$ kicks in for $t\apgt \beta_{typ}^{-1}$ and gives rise to the decay of $S_{EE}(t)$, as in Figs.\ref{fig:Entanglementgrowth}(a),(b), before the system could reach the diagonal ensemble. However, $\beta_{nm}$ has a gap \cite{Can2019}, $\beta_g\sim \mc{O}(1)$, above its minimum value, $\beta_{nm}=0$ (Appendix \ref{sec:Liouvillian_A}). On the contrary, the spectrum of $\delta_{nm}$ is gapless, $\mr{min}(\delta_{nm})\sim  e^{-L}$. Hence, there could be interesting dephasing dynamics that goes on during the decay over $ \beta_{typ}^{-1}\aplt t\aplt \beta_g^{-1}$, even though $\delta_{typ}\gg \beta_{typ}$. For $t>\beta_g^{-1}$, the entanglement entropy rapidly approaches the value for that of the NESS, dictated by the eigenstate with the maximum imaginary part for the eigenvalue, as discussed earlier.

\subsection{Time evolution of N\'{e}el order parameter} \label{subsec:Neel_S} 
We also show the time-evolution of N\'{e}el order parameter $N(t)=(2/L)\sum_i(-1)^i \langle \psi(t)|S_i^z|\psi(t)\rangle$ in Fig.\ref{fig:NeelOrderParameter}. This acts like a \emph{MBL order parameter} \cite{Schreiber2015} and characterizes the memory of the initial state. It approaches a finite value for $t\to\infty$ for the infinite system size in the MBL phase of the Hermitian model; $N(t)$ decays to zero with time in the ergodic phase. Here, for the non-Hermitian case, we find that $N_\infty=N(t\to\infty)$, the N\'{e}el order of the NESS, decreases with $L$, even for strong disorder, deep in the area-law phase. This is expected in a driven system, which loses its initially memory due to the drive. We also notice an interesting dynamical regime, presumably over the time window $\beta_{typ}^{-1}<t<\beta_g^{-1}$, in the decay of $N(t)$ [Fig.\ref{fig:NeelOrderParameter}(b), (c)]. The order parameter plateaus over a region as a function of $t$, as if trying to retain the initial memory. 

In contrast, such regime is absent in the Hermitian model with $\Psi=0$. As shown in Fig.\ref{fig:NeelOrderParameter}(a) for $W=0.43<W_{c1}$, the long-time value $N_\infty$ goes to zero in the ergodic phase. This can be seen from the fact that $N_\infty$ decreases with $L$ tending to zero for $L\to \infty$. On the contrary, in the MBL phase, for $W=4.3,~7$, $N_\infty$ approaches a constant value with increasing $L$. Hence, the N\'{e}el order parameter in this case serves as the MBL order parameter, i.e. it diagnoses the persistence of the initial memory at arbitrary long times. In the non-Hermitian model [Eq.\eqref{eq.SpinModel}], the NESS is approached at long times. This is true even deep in the MBL phase, $W\gg W_{c2}$, since there is always a finite number of eigenstates with complex eigenvalues, albeit with a vanishing fraction, for any finite-size system. Of course, for $W\gg W_{c2}$, it is expected that the imaginary part of the eigenvalues $\sim e^{-L/\zeta}$, and, hence, the NESS will ensue only after a very long time $\sim e^{L/\zeta}$ for large systems. As shown in Fig.\ref{fig:NeelOrderParameter}(b) for $\Psi=0.3$, and in Fig.\ref{fig:NeelOrderParameter}(c) for $\Psi=0.6$, the N\'{e}el order parameter for the NESS decreases rapidly with $L$, presumably approaching zero for $L\to \infty$.

\subsection{Properties of the NESS}
We have so far discussed the entanglement entropy and  current carried by the eigenstates over the entire spectrum of eigenstates. 
In this section we discuss the entanglement entropy and system-size scaling of current for the long-time NESS for time evolution from N\'{e}el state. The NESS is governed by the eigenstate with maximum imaginary part of the eigenvalue or the current [Eq.\eqref{eq:NESS}].  
\subsubsection{Entanglement entropy of NESS}\label{subsec:entanglementNESS_S}
\begin{figure}[!htb]
	\centering{
		\includegraphics[width=\linewidth,height=60mm]{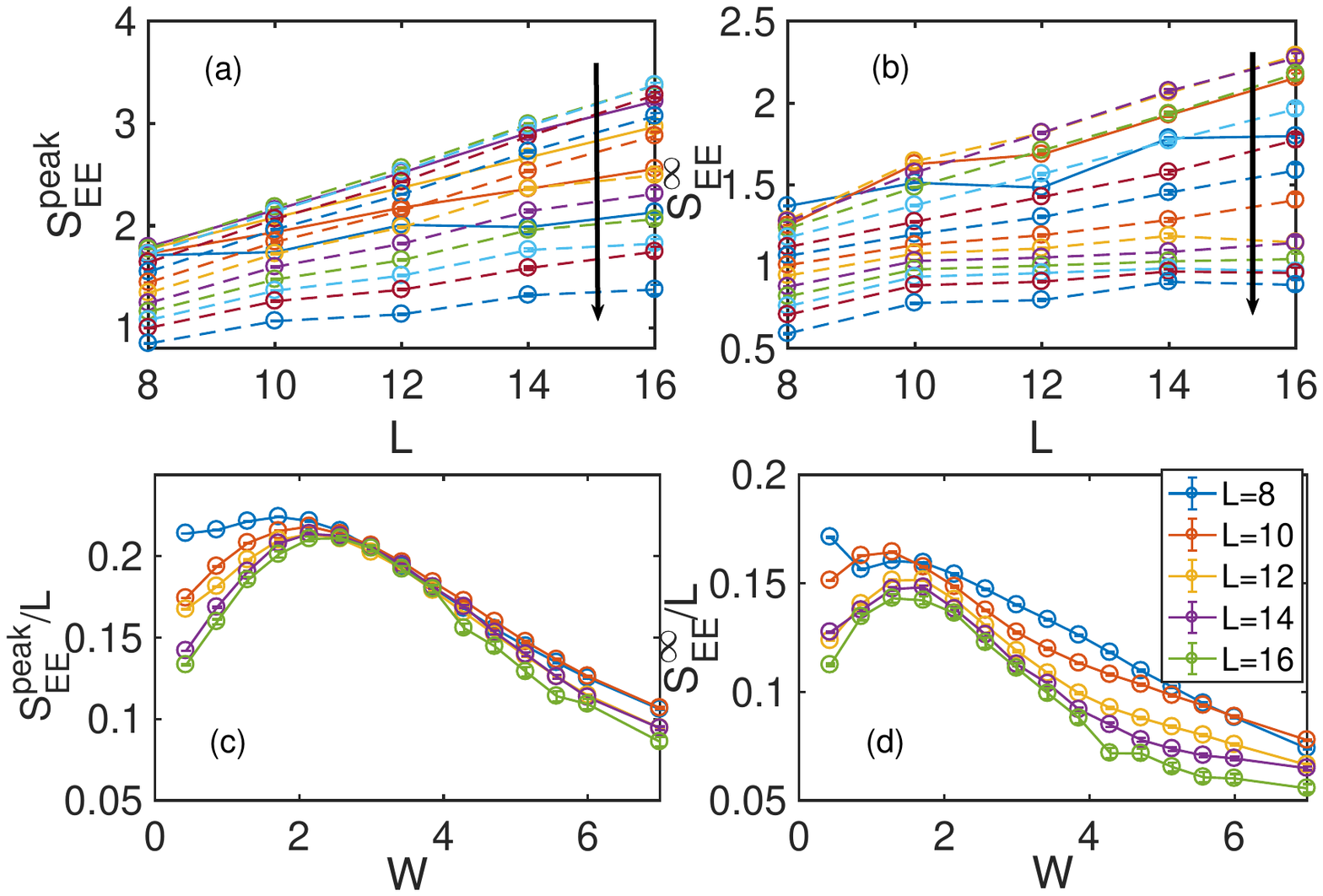}}
	\caption{ \textbf{$S_{EE}^{peak}$ and $S_{EE}^\infty$ for $\Psi=0.3$:} (a) $S_{EE}^{peak}$ and (b) $S_{EE}^\infty$ as a function of $L$ for different $W$. The solid and dashed lines in (a) and (b) are for $W< W_m$ and $W>W_m$, respectively; $W_m=2$ in (a) and $W_m=1.2$ in (b) are the positions of the maxima in $S_{EE}^{peak}$ and $S_{EE}^\infty$, respectively, as function of $W$. In (a) and (b) the arrows indicate the direction of increasing $W$ for the dashed lines and direction of decreasing $W$ for the solid lines. The range of $W$ is from 0.43 to 7. As shown in (c) and (d) unlike average $S_{EE}/L$, $S_{EE}^{peak}/L$ and $S_{EE}^\infty/L$ do not show any crossing when plotted against $W$. These results indicate the absence of any entanglement transition for the NESS.} 
\label{fig:EntanglementPeak}	
\end{figure} 

We denote the entanglement entropy of the NESS by $S_{EE}^\infty$, i.e. $t\to \infty $ limit of $S_{EE}(t)$ [Fig.\ref{fig:Entanglementgrowth}]. We also analyze the maximum value, $S_{EE}^{peak}$, reached by $S_{EE}(t)$ during its time evolution towards NESS. We show $S_{EE}^\infty$ and $S_{EE}^{peak}$ in Figs.\ref{fig:EntanglementPeak}(a), (b), as a function of $W$ and $L$. We find no volume-law to area-law transition in either $S_{EE}^{peak}$ or $S_{EE}^\infty$ with the disorder strength. In fact, $S_{EE}^{peak}$ and $S_{EE}^\infty$ neither scale as the volume nor the area, even though both of these increase with $L$. Hence, the entanglement for the NESSs in the non-Hermitian model of Eq.\eqref{eq.SpinModel} obeys a system-size scaling intermediate between the area- and the volume-law. Similarly, the results for $S_{EE}^{peak}$ indicates that $S_{EE}(t)$ never attains the diagonal ensemble value during its time evolution, unlike that in the Hermitian case \cite{Bardarson2012}. For a given $L$, unlike the eigenstate averaged $S_{EE}$ [Fig.\ref{fig:a2v}(a)], the maxima in $S_{EE}^{peak}$ and $S_{EE}^\infty$ appear at a finite disorder. Also, we do not see any system-size crossing for $S_{EE}^{\infty}/L$ ($S_{EE}^{peak}/L$) vs.~$W$ in Figs.\ref{fig:EntanglementPeak}(c), (d). We were also not able to obtain any finite-size data collapse for $S_{EE}^\infty$ and $S_{EE}^{peak}$.

\subsubsection{NESS current} \label{subsec:Jinf}
\begin{figure}[!htb]
	\centering{
		\includegraphics[width=\linewidth,height=60mm]{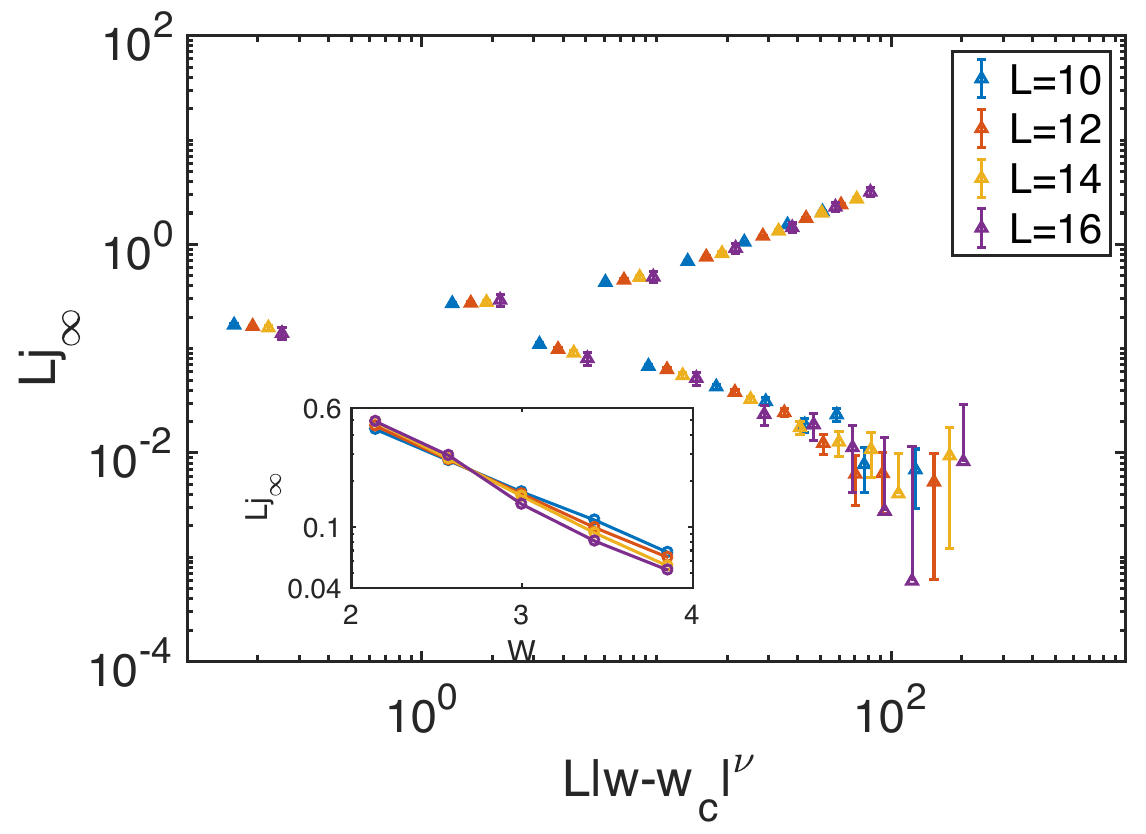}}
	\caption{ \textbf{Current in the NESS:} Finite-size scaling collapse of long-time steady state current $j_\infty$ with $W_c=2.9\pm 0.1$ and $\nu=1.8$ for $\Psi=0.3$. Inset shows the transition in terms of crossing of $Lj_\infty$ vs.~$W$ curves for different $L$, in a semilog plot.} 
	\label{fig:jinf} 
	\FloatBarrier
\end{figure}

There is also a current transition for the NESS, as in Fig.\ref{fig:ScalingCurrent} for the eigenstates. We show the system size scaling of current ($j_\infty$) carried by the NESS in Fig.\ref{fig:jinf}. The transition is shown in terms of a finite-size scaling collapse for $Lj_\infty$ vs.~$W$ curves.

\section{Conclusions and discussions}\label{sec:Conclusions}
 To summarize, we have studied a non-Hermitian disordered model, the interacting version of Hatano-Nelson model \cite{Hatano1996}. We propose that the model can be used to generate a rich variety of current-carrying states and study their entanglement properties. For example, we have found both ergodic and non-ergodic eigenstates with volume and area-law scalings, respectively. Furthermore, we show the existence of a time-reversal symmetry breaking phase transition within the area-law phase. We have established a detailed phase diagram as a function of disorder and strength of the non-Hermitian term based on the properties of the eigenstates and the long-time NESS, as well as, from the time evolution of entanglement entropy.
 
 The models of Eqs.\eqref{eq.SpinModel},\eqref{eq:FermionModel} combine several non-trivial aspects of open quantum systems, namely disorder, interaction and coupling to environment via a non-Hermitian term that also induces a current drive. The disorder tries to localize the spin-flip excitations in the spin model [Eq.\eqref{eq.SpinModel}] or the fermions in the model of Eq.\eqref{eq:FermionModel}. The non-Hermitian term, on the other hand, tries to induce a current and adds to the delocalizing tendency of the hopping term in Eq.\eqref{eq:FermionModel}. This can be easily understood by considering the limit $\Psi \rightarrow \infty$, where the eigenstates of the Hamiltonian are also the eigenstates of the current operator and hence the eigenstates carry finite current. Note that the hopping and the non-Hermitian term in Eq.\eqref{eq:FermionModel} commute with each other. The competition of disorder and current drive and dissipation, already present in the non-interacting Hatano-Nelson model \cite{Hatano1996,Hatano1997,Hatano1998}, leads to a localization-delocalization and $\mc{T}$-reversal breaking transition transition even in 1D. This is in contrast to the Hermitian 1D Anderson model which, as well known, has all the single-particle states localized and there is no transition. 
 
   A similar competition between disorder and non-Hermitian term modeling cooperative decay channel has been studied in other non-Hermitian Anderson models, in the context of collections of overlapping two level systems coupled to light, see, e.g., Refs.\onlinecite{Celardo2014,Biella2013}, which discuss a superradiance (delocalization) to localization transition as a function of disorder. However, the Hatano-Nelson model has the additional aspect of a possible $\mc{T}$-reversal breaking in the eigenstates. 
 
 As in the Hermitian system, the interaction term in Eq.\eqref{eq:FermionModel}, of course, adds to the delocalization tendency by generating matrix elements between exponentially large number of localized many-body product eigenstates of the non-interacting model \cite{Basko2006}. In addition, interaction in the non-Hermitian model separates the localization-delocalization and the $\mc{T}$-reversal breaking transitions, thus leading to the existence of a novel $\mc{T}$-reversal broken area-law phase, unlike that in the non-interacting Hatano-Nelson model. The interaction is also crucial to give rise to intermediate time entanglement growth, absent in the non-interacting model, as shown in Fig.\ref{fig:Entanglementgrowth}. Overall, the non-Hermitian and the interaction terms makes the delocalized phase more robust to disorder.
 
 During the completion of our work we became aware of an independent recent work \cite{Hamazaki2019} which has studied non-Hermitian MBL transition in the same model. Our focus, i.e. to model current driven systems and NESS, is entirely different from that of Ref.\onlinecite{Hamazaki2019}. Our results and conclusions are also substantially different from those in ref.\onlinecite{Hamazaki2019}. In particular, unlike Ref.\onlinecite{Hamazaki2019}, we find that the phase boundaries in $W-\epsilon$ plane for $\mc{T}$ reversal-symmetry breaking and entanglement transitions are distinct. The main reasons behind the difference can be attributed to the choice of parameters, the methodology of detecting phase transition, and, to some extent, interpretation of small-system ED results. The latter is a general problem even for well-studied Hermitian MBL case \cite{Luitz2015,Mace2019}. Ref.\onlinecite{Hamazaki2019} typically focuses on a smaller value of $\Psi=0.1$  and obtain the phase diagram based on level statistics, fraction of imaginary eigenvalues and entanglement entropy averaged over middle one-third of the spectrum, unlike the energy-resolved phase boundary in the  $W-\epsilon$ plane in our case. Due to the smaller value of $\Psi$ the $\mc{T}$ reversal-symmetry breaking and entanglement transitions are much closer, even though different, in Ref.\onlinecite{Hamazaki2019}. Based on some heuristic arguments, the authors of Ref.\onlinecite{Hamazaki2019} conjecture that the small difference between the two transitions will go to zero in the thermodynamic limit. On the contrary, our results with detailed finite-size scaling analysis in the $W-\epsilon$ plane shows that the two phase boundaries are clearly distinct for the particular choice of parameters ($\Psi=0.3,0.6$) in our work. We also note that our results for the ergodic to non-ergodic transition is consistent with a more recent work \cite{White2020} on the same model, studied in a different context. Moreover, we give detailed insight into the approach to NESS and identify new dynamical regime where memory of initial state can be harnessed over a relatively long intermediate time window during the relaxation to the NESS (Sections \ref{subsec:Neel_S}).

In future, it would be also interesting to understand whether the non-Hermitian model can indeed describe some features of a many-body localizable system under an actual current drive or voltage bias applied through leads.

{\bf Acknowledgement:} We thank Subroto Mukerjee, Abhishek Dhar and Adhip Agarwala for useful discussions. We also thank Marko Znidaric and Nicolas Laflorencie for their valuable comments on our manuscript. SB acknowledges support from The Infosys Foundation (India), and SERB (DST, India) ECR award  . 

\appendix

\section{Spectrum of the Liouvillian operator } \label{sec:Liouvillian_A}
\begin{figure}
	\centering{
		\includegraphics[width=\linewidth,height=60mm]{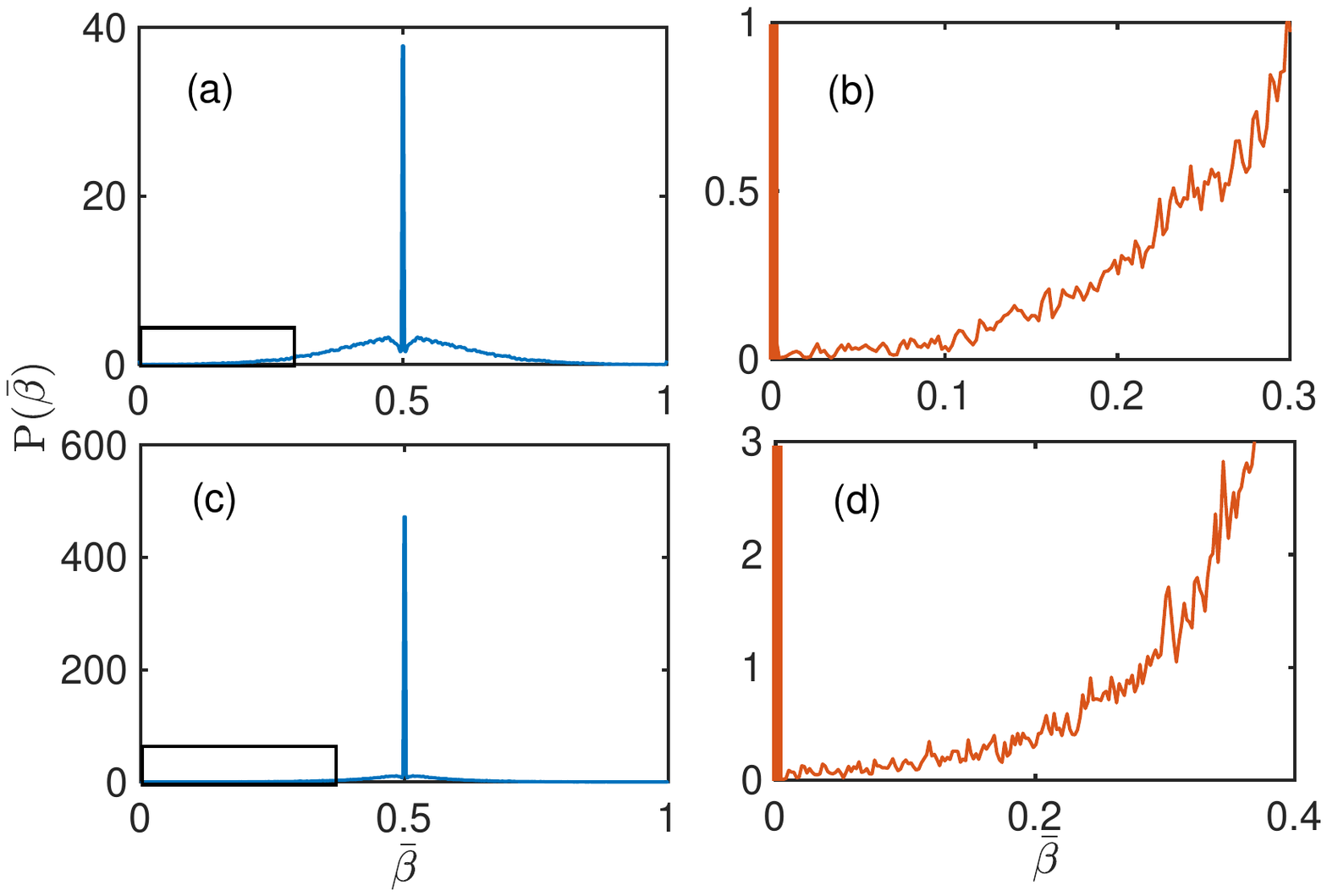}}
	\caption{\textbf{Probability distribution of the real part of the eigenvalues of the Liouvillian operator:} The probability distribution P($\bar{\beta}$) for (a) $W = 1$ and (b) $W = 2$ for $\Psi=0.3$. The boxes in (a) and (c) highlight the gapped part	of the spectra, which are zoomed in (b) and (d), respectively.}
	\label{fig:Liouvillian_A} 
\end{figure}
As discussed in the main text, the time evolution of the density matrix is controlled by the eigenspectrum of the Liouvillian operator $\mc{L}$ in Eq.\eqref{eq.DynEq}. The relaxation of the system to the NESS is controlled by the real part of the eigenvalues of  $\mc{L}$, i.e. $-\beta_{nm}>0$, where $\beta_{nm}$ $=2\Lambda_s-\Lambda_n-\Lambda_m$. In Figs.\ref{fig:Liouvillian_A}, we plot the disorder averaged distribution, $P(\bar{\beta})$, of the normalized quantity, $\bar{\beta}=(\beta-\beta_0)/(\beta_M-\beta_0)$ for $\Psi=0.3$. Here $\beta_0$ and $\beta_M$ are the minimum and the maximum values of $\beta_{nm}$, respectively. Since the imaginary eigenvalues of the non-Hermitian Hamiltonian in Eq.\eqref{eq.SpinModel} appears in complex conjugate pairs, the probability distribution is symmetric about $\bar{\beta}=1/2$. Also, there is a peak at $\bar{\beta}=0 $, which corresponds to the NESS, as mentioned in the main text. As shown in Figs.\ref{fig:Liouvillian_A}, the peak is separated from most of the spectrum by a long tail. In the limit $L\to\infty$, the tail is expected to tend to a gap \cite{Can2019}, $\tilde{\beta}_g$, that separates the peak from the relaxation modes with $\bar{\beta}>0$.

\vspace{1mm}
\bibliography{MBL}

\end{document}